\documentclass[aps,twocolumn,showpacs,preprintnumbers,amsmath,amssymb,nofootinbi
b,superscriptaddress,showkeys]{revtex4}

\usepackage{epsfig}
\usepackage{graphicx}

\begin{document}

\title{Renormalization approach to constituent quark models of
  {\bf Q}uarkonium\footnote{Supported by Ministerio de Ciencia y
    Tecnolog\'\i a under Contracts no. FPA2010-21750-C02-02, by the
    European Community-Research Infrastructure Integrating Activity ''Study of
    Strongly Interacting Matter'' (HadronPhysics2 Grant no. 227431),
    the Spanish Ingenio-Consolider 2010 Program CPAN (CSD2007-00042),
    the Spanish DGI and FEDER funds with grant no. FIS2008-01143/FIS,
    Junta de Andaluc{\'\i}a grants no. FQM225.}}
\author{J. Segovia}\email{segonza@usal.es}
\author{D.R. Entem}\email{entem@usal.es}
\author{F. Fern\'andez}\email{fdz@usal.es}
\affiliation{Grupo de F\'{\i}sica Nuclear and IUFFyM, \\ Universidad de Salamanca, 
E-37008 Salamanca, Spain}
\author{E. Ruiz Arriola}\email{earriola@ugr.es} \affiliation{Departamento de 
F\'isica At\'omica, Molecular y Nuclear, \\ Universidad de Granada, E-18071 
Granada, Spain.}
\affiliation{ Instituto Carlos I de F{\'\i}sica Te\'orica y Computacional,
  Universidad de Granada, E-18071 Granada, Spain}
\date{\today}

\begin{abstract} 
\rule{0ex}{3ex} Constituent quark models, while successful, require a
great deal of fine tuning of the short distance interactions by
introducing phenomenological gluonic form factors which are ultimately
designed to accurately reproduce the spectrum. We apply and develop
renormalization ideas to reduce the short distance sensitivity and
show that, as naively expected, but not explicitly implemented in the
models, the physics of binding is entirely linked to the string
tension whereas leptonic decays depend more on the gluon exchange
potential.  We also show how the spectrum of S- and D-
$1^{--}$ states is successfully intertwined through the singular tensor 
interaction. 

\end{abstract}


\maketitle



\section{Introduction}

After the discovery of the first heavy-quark bound states, the $\psi$
and the $\Upsilon$ systems, it was soon realized that a
nonrelativistic picture seemed to hold for them. The charmonium system
has become the prototypical 'hydrogen atom' of meson spectroscopy and
therefore it was the first and simplest case where a bound state of
quarks could be studied. The old Cornell
potential~\cite{Eichten:1979ms}
\begin{eqnarray}
V(r)=-\frac{4\alpha_{s}}{3r}+\sigma r
\end{eqnarray}
provides a rough estimate of the  $1^{--}$ charmonium spectrum and in particular 
of the $J/\psi$ resonance (see e.g.
Ref.~\cite{Lichtenberg:1987ms,Voloshin:2007dx} for a review and references
therein). In potential models, One-Gluon Exchange (OGE) short distance
singularities appear and phenomenological gluonic form factors must be
introduced (see e.g.~\cite{Weinstein:1983gd} for an early proposal).  This
triggers an unpleasant short distance sensitivity, and as we will show it mainly
hides the fact that the ground state is actually being used as an input rather
than a prediction. In this paper we want to make this statement more
quantitative and to analyze to what extent can one disentangle the physics of
the ground state to that of the excited states. The method which we will be
using is based on renormalization ideas. To provide a proper and broad
perspective and to avoid confusion with other approaches closely related, it is
of interest to review some relevant developments in what follows. 

For the lowest-lying bound states the size of the system is smaller or
comparable than $1/\Lambda_{\rm QCD}$~\cite{Titard:1993nn,Titard:1994id},
whereas excited states start feeling the long range string tension. In the case
of charmonium this is so until one reaches the $\bar{D}D$ production
threshold. Of course, one expects that at short distances corrections are
computable within perturbation theory due to asymptotic freedom, whereas
at long distances one  must resort to lattice calculations.

For heavy particles the static energy, $E(R)- 2 m $, is an observable
and thus a gauge and scale independent quantity. The rigorous
non-perturbative definition of the $\bar Q Q$ potential is the
expectation value of the Wilson loop and has been continuously used on
lattice calculations~(a review on potentials from lattice QCD can be
found e.g. in Ref.~\cite{Bali:2000gf}). It should be noticed that
potentials are computed for point-like sources, i.e. structureless
quarks. The determination of the $\bar Q Q$ potential has been carried
out in perturbation theory up to $N^3
LO$~\cite{PhysRevLett.104.112002,PhysRevLett.104.112003}. Looking for
a scale independent potential generates the Renormalon problem (for a
review see e.g. Ref.~\cite{Beneke:1998ui}) which causes serious
convergence difficulties.  Effective Field Theory (EFT) methods
disentangle the hard, soft and ultrasoft scales
perturbatively~\cite{Brambilla:2004jw} and explicitly solve the
problem.

The pNRQCD approach~\cite{Pineda:1997bj,Brambilla:1999xf} is an EFT which has
better convergence properties than NRQCD since it encodes the ultrasoft scales
non-perturbatively in terms of potentials which enter as Wilson
coefficients and should be obtained directly from the lattice or string models
enabling a reliable description of excited states. The bottomline seems to be
that perturbative potentials should be treated perturbatively whereas non
perturbative potentials should be treated non-perturbatively.

Following an observation~\cite{Beneke:1998rk} that the renormalon problem is
triggered by an illegal extension of Fourier integrals to all momenta, a
parametrization of the lattice data in the short distance  region, $r \le
0.15 {\rm fm}$, has been proposed~\cite{Laschka:2011zr} cutting off low momenta
below $\mu_f \sim 1 {\rm GeV}$. If $\mu_f \to 0$ a flagrant lack of convergence
of perturbation theory has been pictorially represented in 
Ref.~\cite{Laschka:2011zr};  depending on the scheme proposed one may
even violate fundamental inequalities~\cite{Bachas:1985xs,Nussinov:2000kn}. It
should be noted that this representation smoothly matches the current lattice
determination of the static energy allowing an analysis of the quarkonium
spectrum, but suggests that the non-perturbative contribution corresponds to $q
\le \mu_f$, larger than the ultrasoft scales expected from pNRQCD. Unlike the
$\mu_f \to 0$ case the convergence is very good and basically a simple OGE
potential is left.

Of course all these considerations could less transparently be handled from a
fundamental point of view by solving the theory {\it ab initio}. However, direct
lattice QCD calculations of quarkonium spectrum require small enough lattice
spacings, so that $M a \ll 1 $ and so far promising calculations operate at $a
\sim 0.1-0.2 {\rm fm}$ producing some over-estimation of excited
states~\cite{Burch:2009az}.

Turning to a less fundamental approach such as the non-relativistic
quark model and in the spirit of the Cornell potential and extensions
of it let us mention that the standard regularization of the short
distance contributions to the OGE potential (corresponding to delta
contact interactions and the tensor and spin-orbit pieces), actually
corresponds to tune the parameters of the regulator to fit the
$J/\psi$ mass. In addition, one of the usual parameters which are
adjusted is the strong coupling constant, $\alpha_{s}$. However, the
world average strong coupling constant is
$\alpha_{s}(M_{Z^{0}})=0.1184\pm0.0007$~\cite{Bethke:2009jm}. The
measured values of $\alpha_{s}(Q^{2})$, covering energy scales from
$Q\equiv M_{\tau}=1.78 {\rm GeV}$ to $209 {\rm GeV}$, exactly follow
the energy dependence predicted by QCD and therefore significantly
test the concept of Asymptotic Freedom. Therefore, the value of
$\alpha_{s}$ at the charm quark mass is fixed and is not larger than
$0.2-0.3$. This value contradicts some quark models.

A nonperturbative determination of the QCD potential at ${\cal O}(1/m)$ was
undertaken for the first time in~\cite{Koma:2006si,Koma:2006fw}, they found a
potential which depends on the interquark distance as $-c/r^{2}$ and it is
comparable with the coulombic term of the static potential when applied to
charmonium and amounts to one-fourth of the coulombic term for bottomonium.

Lattice calculations of $1/m$ corrections based on the Bethe-Salpeter
amplitude~\cite{Ikeda:2011bs} (see also \cite{Kawanai:2011xb}) reveal
that the dependence with the interquark distance of the $1/m$
contributions to the potential is the same than in the case above but
other term which grows logarithmically with $r$ was found. When the
data are fitted taking into account LO and NLO contributions to the
potential, the coefficients which accompany Coulomb and string forces
are comparable to those calculated through Wilson loops.

The ${\cal O}(1/m^2)$ spin-dependent and momentum-dependent corrections are
investigated in SU(3) lattice gauge theory~\cite{Koma:2009ws}. These
corrections are relevant ingredients of an effective field theory for heavy
quarkonium called potential nonrelativistic QCD.

In the present paper we want to address the question on how much can
we deduce for the charmonium spectrum from the knowledge of the
potential at long distances in a way that our ignorance at short
distances needs not play a crucial role. This allows to disentangle
the physics of the ground state to that of the excited states which
are sensitive to the perturbative (Coulomb like) and non-perturbative
(String like) corrections. Actually, our original motivation for the
present study was to analyze the role of regulators within constituent
quark models like that of Ref.~\cite{Vijande:2005eu}. However, our
ideas can be generally extended to any quark model and shed some light
on their predictive power.

The plan of the paper is as follows. In Section~\ref{sec:toy} we
provide a simplified and pedagogical discussion of our main points in
the simplest case of the Cornell potential. The role of singular
tensor and spin-orbit interactions is analyzed in
Section~\ref{sec:toy-coupled} where an interesting correlation between
S- and D-waves is found. In Section~\ref{sec:oldmodel} we review the
bases of the constituent quark model proposed in
Ref.~\cite{Vijande:2005eu}. In Section~\ref{sec:renormodel} we focus
on the new renormalized charmonium model where the short distance
regulators have been removed as they turn out to be physically
irrelevant for most observables. Interesting issues regarding the scope and applicability
of these renormalization ideas are scrutinized in
Section~\ref{sec:applic}. Further aspects are dealt with by an
enlightening analysis of the Bosonic String Model (BSM),
Section~\ref{sec:bsm}.  Finally, we give some conclusions in
Section~\ref{sec:conclusions}.

\section{Renormalization of the Cornell Potential}
\label{sec:toy}

In this section, we provide a comprehensive discussion of the
renormalization approach as applied to heavy quark systems and  
confining potentials within the context of non-relativistic quantum
mechanics. The Schr\"odinger equation is given by
\begin{eqnarray}
-\frac{1}{2 \mu_{\bar q q }} \nabla^2 \Psi  + V(r)  \, \Psi &=&  \, 
(M-m_q - m_{\bar q}) \Psi,
\end{eqnarray}
where $\mu_{\bar q q }= m_q m_{\bar q}/(m_q + m_{\bar q}) $ is the
reduced $\bar q-q$ mass and the normalization condition $\int d^3 r
|\Psi(\vec r)|^2$ = 1 must be imposed on the solution. We use the
standard Cornell potential  
\begin{eqnarray}
V(r)=-\frac{4\alpha_{s}}{3r}+\sigma r,
\label{eq:pot-cornell}
\end{eqnarray}
for the purposes of illustration. The more elaborated potentials,
where relativistic effects and spin-dependent corrections are added, 
will be discussed in later Sections. It is remarkable that such a
simple potential not only captures the relevant physics of the problem
but also is accurately described by lattice calculations (see
e.g.~\cite{Necco:2001xg}) where Eq.~(\ref{eq:pot-cornell}) is favoured
with $4 \alpha_s/3= 0.25(1)$, extremely close to the bosonic string
model $\alpha_s=\pi/16$~\cite{Alvarez:1981kc,Arvis:1983fp}. It is worth
mentioning that the lattice calculations of $\bar q - q$ potentials
correspond to using point sources, i.e., elementary quarks until
distances comparable to the lattice spacing. A smooth transition below
$a \sim 0.2 \,{\rm fm}$ towards OGE has been observed. Unless otherwise stated
we will take $\alpha_s= \pi/16$, $\sigma= (420 \, {\rm MeV})^2=0.1764 \, 
{\rm GeV}^2$, $M_{J/\psi}=3096.916 \, {\rm MeV}$, 
$m_c=1200\,{\rm MeV}$ and $r_c=0.3-0.01 \, {\rm fm}$. 

For the spherically symmetric Cornell potential the total relative
wave function can be factorized in the usual fashion, $\Psi(\vec r) =
(u(r)/ r) Y_{l m} (\hat r)$, with $u(r)$ the reduced wave function and
$Y_{lm}(\hat r)$ the conventional spherical harmonics.

\subsection{Review of the Renormalization approach}

Let us consider the standard non-relativistic Schr\"odinger equation for bound 
states in S-waves 
\begin{eqnarray}
-\frac{1}{2\mu}u_{n}''(r)+V(r)u_{n}(r)=E_{n}u_{n}(r),
\label{eq:sch_Swave}
\end{eqnarray}
where $u_{n}(r)$ vanishes at long distances and the energy is defined
with respect to the $\bar q - q$ threshold, $E_n = M_n - m_q - m_{\bar q}$. 

Anticipating our discussion we will assume for definiteness a short
distance auxiliary cut-off, $r_{c}$, below which the potential
vanishes. This cut-off is just a parameter which will ultimately be
removed while keeping some physical condition fixed. In our case we
will choose to fix the ground state energy to the experimental
value. In the numerical application we will be concerned with the
residual cut-off dependence of observables induced by such a
procedure.

Using the standard trick of multiplying the Eq.~(\ref{eq:sch_Swave}) by
$u_{m}(r)$ and subtracting the similar equation with $(n \leftrightarrow m)$, we
get for two different energies $E_{n}\neq E_{m}$ the orthogonality relation
between their bound state wave functions
\begin{eqnarray}
&& u_{n}' (r_{c}) u_{m} (r_{c}) - u_{n}(r_{c}) u_{m}' (r_{c}) \nonumber \\ &&= 2
  \mu(E_{n}-E_{m})\int_{r_{c}}^\infty u_{n}(r) u_{m}(r) \, d r.
\end{eqnarray}

Note that {\it usually} the regularity condition at the origin, 
$u_{n}(r_{c})=0$ for $r_{c}\to0$, is imposed.  Whence orthogonality of wave 
functions with different energies holds.  However, this is not the only 
solution to the orthogonality requirement. Instead the 
common boundary condition, we may as well take
\begin{eqnarray}
\frac{u_{m}'(r_{c})}{u_{m}(r_{c})}=\frac{u_{n}'(r_{c})}{u_{n}(r_{c})},
\end{eqnarray}
for any two states, meaning that the logarithmic derivative at short distances 
becomes state independent. In particular, choosing the ground state as a 
reference state we get the condition
\begin{eqnarray}
\frac{u_{0}'(r_{c})}{u_{0}(r_{c})}=\frac{u_{n}'(r_{c})}{u_{n}(r_{c})}.
\label{eq:bc}
\end{eqnarray}

How can this logarithmic derivative be determined? If we know the
energy of the ground state and the potential we may integrate from the
long distance region inward to deduce $u_{0}'(r_{c})/u_{0}(r_{c})$.
Once this number is known, we may use Eq.~(\ref{eq:bc}) to integrate
out the excited state and the corresponding bound state energy can be
fixed by requiring the wave function to vanish at large
distances. Thus, such a procedure allows to treat the ground state
energy, $E_{0}$, as an independent variable from the potential $V(r)$
and still deduce wave functions and the excited spectrum.  There is of
course the question on how to interpret the short distance cut-off,
$r_{c}$. In principle one may look for stability at scales below the
relevant sizes.  Actually, by varying the cut-off in this region this
is a way of assessing theoretical uncertainties~\footnote{In a model
where phenomenological form factors are implemented {\it ad hoc}, just
to prevent singularities, the assessment of theoretical errors could
be done by choosing {\it all possible} regularization functions,
unless the form factor is known from first principles.}. However,
there are situations where the limit $r_{c}\to0$ is rather smooth and
induces moderate changes in observables.

A good feature of the present approach is that since wave functions are matched
at short distances the resulting energies are largely independent on the short
distance behaviour of the potential~\footnote{The previous approach is well
documented in the mathematical literature as the theory of self-adjoint
extensions of hermitian operators. This endows the Hilbert space with a
common domain of functions where the solutions of the Schr\"odinger equation
span a complete set. The generation of a new scale independently on the
potential resembles the well known phenomenon of dimensional transmutation.}.

The previous discussion has been conducted for $S$-waves and regular
potentials, i.e. fulfilling $ \lim_{r \to 0} r^2 |V(r)| <
\infty$. Higher partial waves cannot be renormalized in this fashion
as short distances are dominated by the centrifugal barrier. This also
has the benefit of diminishing the dependence on the short distance
potential since the short distance behaviour of the wave function,
$u(r) \sim r^{l+1}$, is not controlled by the potential.

The case of singular potentials has also been discussed at length (see
e.g. Refs.~\cite{RuizArriola:2007wm,PavonValderrama:2005wv,Arriola:2009wi}
regarding nuclear or atomic systems). The relation to momentum space
renormalization and the corresponding Lagrangian counterterms is discussed in
Ref.~\cite{Entem:2007jg}. The irrelevance of form factors is analyzed in
Ref.~\cite{Arriola:2009wi,Cordon:2009pj}.

\subsection{Analysis of the linear potential}

In order to provide a clear picture of the procedure to be carried out
in the present work, we discuss the issue within the framework of the
simplified model where the OGE piece is neglected. For $S$-wave
charmonium states we have the simplified problem
\begin{eqnarray}
-\frac{1}{m_c} u_{c\bar{c}}''(r)+\sigma r  \, u_{c\bar{c}}(r) &=& E \, 
u_{c\bar{c}} (r). 
\end{eqnarray}
As it is well known, the general solution in the inner region is a
linear combination of regular and irregular Airy functions
\begin{eqnarray}
u_{c\bar{c}}(r)=c_{1}{\rm Ai}(z)+c_{2}{\rm Bi}(z),
\end{eqnarray}
where the dimensionless variable 
\begin{eqnarray}
z=\frac{m_{c}(-M+2m_{c}+r \sigma)}{(m_{c}\,\sigma)^{2/3}}
\label{eq:z_def}
\end{eqnarray}
has been introduced. At large values of the argument one has
\begin{eqnarray}
{\rm Ai}(z) &=&  \frac{e^{-\frac{2}{3}z^{3/2}}}{2 \sqrt{\pi} z^{1/4}}\left[ 1 + 
{\cal O}(z^{-3/2}) \right], \nonumber \\
{\rm Bi}(z) &=&  \frac{e^{\frac{2}{3} z^{3/2}}}{\sqrt{\pi} z^{1/4}}\left[ 1 + 
{\cal O}(z^{-3/2}) \right], 
\end{eqnarray}
so that for bound states and assuming no opening of meson decay channels we may 
discard the ${\rm Bi}(z)$ function. The standard approach consists of 
requiring the regularity condition at the origin.

Indeed, the regular solution at the origin requires finding the zeros of the Airy 
function, ${\rm Ai} (z_n)=0$, yielding the quantization formula
\begin{eqnarray}
z_{n}=\frac{m_{c}(-M_{n}+2 m_{c})}{(m_{c}\,\sigma)^{2/3}},
\end{eqnarray}
where $z_{0}=-2.33811,z_{1}=-4.08795,z_{2}=-5.52056,z_{3}=-6.78671,
z_{4}=-7.94413,\dots$ are the corresponding lowest zeros.  Note that
in this particular case the string tension and the quark mass
determine the binding energy completely.

\begin{figure}[ttt]
\begin{center}
\epsfig{figure=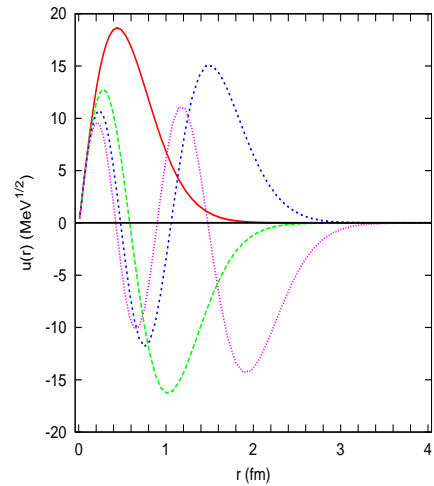,height=6.5cm,width=6cm}
\end{center}
\caption{Wave functions for the different states. The normalization is
  such that at the matching point, $r_{c} \to 0 {\rm fm}$, the
  functions and derivatives of the different states coincide.}
\label{fig:wave-functions}
\end{figure}

An alternative procedure might be as follows. Since we want to fit the
$J/\psi$ mass and we are confident that the long distance
dynamics is given by the linear potential~\cite{Bali:2000gf}, we are
quite certain that at long distances the wave function is given by
\begin{eqnarray}
u_{c\bar{c}}(r)=c_{1}{\rm Ai}(z),
\label{eq:asympt}
\end{eqnarray}
where now $M_{J/\psi}$ is taken as an {\it input}. Of course, for a
generic value of the string tension the wave function will fail to
vanish at the origin (see Fig.~\ref{fig:wave-functions}). This may be
seen as a drawback but the gain is in the {\it prediction} of the
excited spectrum. This is done by matching logarithmic derivatives at
the origin or at a short distance cut-off, Eq.~(\ref{eq:bc}), yielding
\begin{eqnarray}
\frac{{\rm Ai}'(z_{0})}{{\rm Ai}(z_{0})}=\frac{{\rm Ai}'(z_{n})}{{\rm 
Ai}(z_{n})},
\end{eqnarray}
where $z_{n}$ is given by 
\begin{eqnarray}
z_{n}=\frac{m_{c}(-M_{n}+2m_{c}+\sigma r_{c})}{(m_{c}\sigma)^{2/3}}.
\end{eqnarray}

\begin{figure}[ttt]
\begin{center}
\epsfig{figure=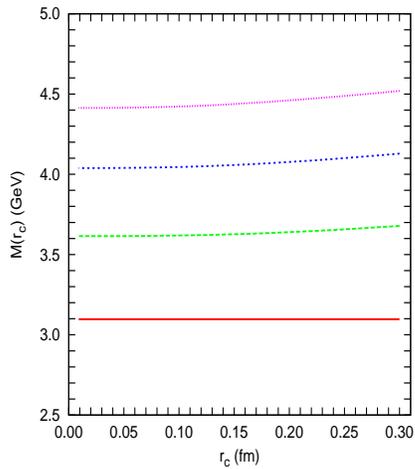,height=6.5cm,width=6cm}
\end{center}
\caption{Dependence of the excited charmonium states on the short distance 
cut-off.}
\label{fig:cut-off}
\end{figure}

This condition guarantees the orthogonality of states and provides
$M_{n}$ from $M_{0}$ for any value of $r_{c}$. Note that by definition
the short distance behaviour of the wave functions is very similar, so
that even though we may not know accurately the potential at short
distances there is an increasingly large cancellation. The cut-off
dependence of the states is depicted in
Fig.~\ref{fig:cut-off}~\footnote{The cut-off dependence can be
understood on purely analytical grounds as carried out in
Appendix~\ref{sec:errors} for a wide class of potentials.}. As we see,
changing the scale provides a mild dependence featuring the advertised
short distance insensitivity.

\subsection{Inclusion of One-Gluon Exchange}

In heavy quark systems the OGE potential is considered to be a very
important ingredient to provide some additional short range
corrections. Here we will show that as far as binding is concerned
this OGE contribution becomes almost irrelevant. Contrary to what one
might naively think this is actually good news.  Indeed, as we
mentioned in the introduction the short distance part of the potential
is not well known even at the OGE level. If we write the OGE potential
in the form $V(r)=-4\alpha_s(\mu)/3r$, with $\mu$ the $\overline{MS}$
renormalization scale, it is not obvious what scale should one use
{\it a priori}. Lattice calculations display such a behaviour at
short distances (see the recent update where running is actually
observed~\cite{Donnellan:2010mx}). There have been attempts to improve
on this by imposing renormalization group invariance of the
potential. Unfortunately such a procedure breaks down due to the
appearance of renormalons (see e.g. Ref.~\cite{Beneke:1998ui} and
\cite{Laschka:2011zr}) which spoil a convergence pattern. In addition,
there is also the problem on how far should this OGE force be
extended, since it only applies to very short distances. However, one
expects that as compared to the linear potential the long range effect
is not crucial.

\begin{figure}[ttt]
\begin{center}
\epsfig{figure=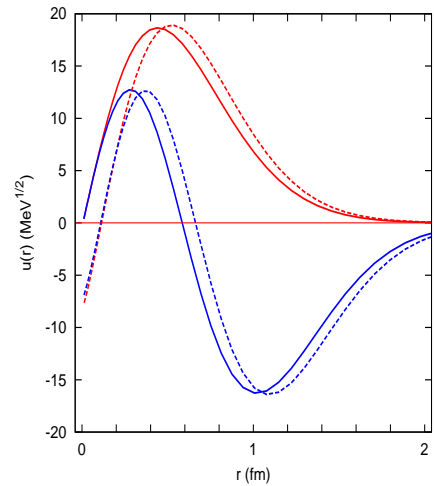,height=6.5cm,width=6cm}
\end{center}
\caption{Charmonium wave functions for $\psi(1S)$ and $\psi(2S)$
states without (solid line) and with (dashed line) One Gluon Exchange potential. 
}
\label{fig:wave-functions-OGE}
\end{figure}

The region where OGE becomes comparable with the linear potential is $r\sim
b=\sqrt{\alpha_s/\sigma} \sim 0.2-0.3 {\rm fm}$ where the total potential is
about $V(r) \sim -180 {\rm MeV}$. At much larger distances we may neglect the
OGE component so that we may take the asymptotic behaviour of
Eq.~(\ref{eq:asympt}). The resulting wave functions without and with OGE are
depicted in Fig.~\ref{fig:wave-functions-OGE}. Similarly to the case without OGE
the procedure is as follows. We start from the {\it given} ground state energy
$M_0= M_{J/\psi}$ integrating inward down to the short distance cut-off radius,
$r_c$, the full Cornell potential, Eq.~(\ref{eq:pot-cornell}). To facilitate the
comparison we take the same asymptotic wave function for both cases,
without or with OGE. As can be seen, the OGE attraction provides a larger
probability in the inner region below the scale where OGE starts being
negligible. Our numerical calculations indicate that this long distance scale is
about $1.5 {\rm fm}$. From the ground state we can construct the remaining
excited states by matching logarithmic derivatives at the short distance
cut-off, as indicated above.

\begin{figure}[ttt]
\begin{center}
\epsfig{figure=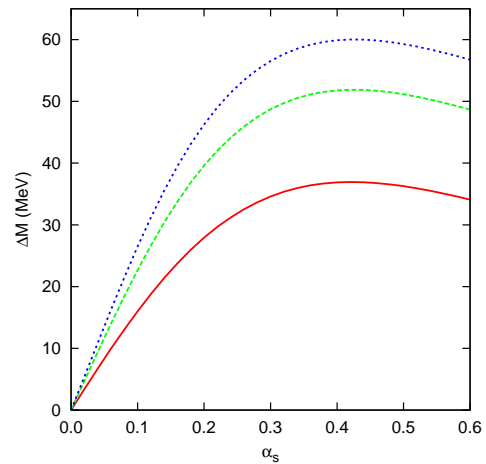,height=6.5cm,width=7cm}
\end{center}
\caption{Mass shift (in MeV) as a function of the strong coupling
  constant $\alpha_s$ when the short distance cut-off is removed $r_c
  \to 0$ for the $\psi(2S)$, $\psi(3S)$ and $\psi(4S)$ relative to the
  ground $\psi(1S)$ state.}
\label{fig:mass-shift-OGE}
\end{figure}

The short distance cut-off dependence of the excited charmonium states for
several values of the strong coupling constant is similar in all cases to the
$\alpha_{s}=0$ situation shown in Fig.~\ref{fig:cut-off}. While at large $r_{c}$
the calculation is dominated by the linear potential, wave functions at short
distances are very much alike due to the common boundary condition. Thus, the
difference in energy due to OGE comes from the energy dependence of the wave
function below the non-perturbative scale. In Table~\ref{tab:masses-toy},
for the case $r_c \to 0$, we show  numerical values of the excited states
for $\alpha_{s}=0.0,\,0.1,\,0.2\mbox{ and }0.3$. The mass shift $\Delta M_n =
M_n-M_0$ is shown in Fig.~\ref{fig:mass-shift-OGE} for $r_c = 0.06 {\rm fm}$
and as we can see the OGE effects are rather moderate.

\begin{table}[t!]
\begin{center}
\begin{tabular}{ccccccccc}
\hline
\hline
$n$ & & $\alpha_{s}=0.0$ & & $\alpha_{s}=0.1$ & & $\alpha_{s}=0.2$ & &
$\alpha_{s}=0.3$ \\[2ex]
$1$ & & $3615$ & & $3630$ & & $3638$ & & $3640$ \\
$2$ & & $4039$ & & $4060$ & & $4070$ & & $4073$ \\
$3$ & & $4414$ & & $4439$ & & $4449$ & & $4452$ \\
$4$ & & $4756$ & & $4783$ & & $4795$ & & $4798$ \\
\hline
\hline
\end{tabular}
\caption{\label{tab:masses-toy} Masses in MeV of S-wave charmonium states
for several values of the $\alpha_{s}$ coupling constant and keeping always the
ground state mass $M_{0}=3096.916\,\mbox{MeV}$.}
\end{center}
\end{table}

\section{Coupled channel solutions: Mixing S and D states}
\label{sec:toy-coupled}

In this section we include the spin dependent potential contributions
coming from the OGE interaction to the simple Cornell potential discussed in
Section~\ref{sec:toy}.

\subsection{Potential}

The $\bar{q}q$ interaction can be written as 
\begin{equation}
V (\vec r)  = \sigma r + V_{\rm OGE} ( \vec r ),
\end{equation}
where now $V_{\rm OGE}(\vec{r})$ contains central, the coulomb interaction, and
non-central, tensor and spin-orbit, contributions 
\begin{eqnarray}
V_{OGE}^{C}(\vec{r}) &=&-\frac{4\alpha_{s}}{3}\frac{1}{r}, \\ 
V_{OGE}^{T}(\vec{r}) &=& \frac{1}{3}\frac{\alpha_{s}}{m^2}\frac{1}{r^3} S_{12}, \\
V_{OGE}^{SO}(\vec{r}) &=& \frac{2 \alpha_{s}}{m^2}\frac{1}{r^{3}}  \vec{L} \cdot \vec{S} \, ,  
\label{eq:OGE_unreg}
\end{eqnarray}
where $\alpha_{s}$ is the strong coupling constant, $m$ is the equal
quark and antiquark mass,
$S_{12}=3\,(\vec{\sigma}_{1}\cdot\hat{r})(\vec{\sigma}_{2}
\cdot\hat{r} )-\vec{\sigma}_{1} \cdot\vec{\sigma}_{2}$ is the quark
tensor operator, $\vec{S} $ the total spin operator and $\vec{L} $ is
the relative orbital angular momentum operator.

\subsection{The coupled S-D equations}

Our previous discussion of renormalization was undertaken without
taking into account the role played by the tensor and spin-orbit
forces. In the case of charmonium, these states are a combination of S
and D-wave components due to the tensor force contribution. As we will
see below, this tensor force is small enough to have almost pure S and
D orbital state components. The interesting aspect of our discussion
below is that using just one renormalization condition we can predict
all S and D-wave mesons, i.e. we reduce the number of renormalization
conditions.  

The radial Schr\"odinger equation for the $^{3}\!S_{1}-\,^{3}\!D_{1}$ coupled
channel reads 
\begin{equation}
\begin{split}
& -u''(r)+U_{S}(r)u(r)+U_{SD}(r)w(r)= \kappa^2 u(r), \\
& -w''(r)+U_{SD}(r)u(r)+\left[U_{D}(r)+\frac{6}{r^{2}}\right]w(r)= \kappa^2 w(r),
\label{ref:CE}
\end{split}
\end{equation}
where $\kappa^2 =m_c (M- 2 m_c)$. $U_{S}$, $U_{D}$
and $U_{SD}$ are the different contributions of the reduced potential,
$U(r)=2\mu V(r)$, where the $^{3}\!S_{1}-\,^{3}\!D_{1}$ coupled
channel potential is given by
\begin{equation}
\begin{split}
&
V_{S}(r)=-\frac{4\alpha_{s}}{3r}+\sigma r, \\
&
V_{D}(r)=-\frac{4\alpha_{s}}{3r}+\sigma
r-\frac{20}{3}\frac{\alpha_{s}}{m_{c}^{2}}\frac{1}{r^{3}}, \\
&
V_{SD}(r)=\frac{2\sqrt{2}}{3}\frac{\alpha_{s}}{m_{c}^{2}}\frac{1}{r^{3}},
\end{split}
\end{equation}
in which $\alpha_{s}$ is the strong coupling constant, $\sigma$ is the string
tension and $m_{c}$ is the quark mass. Obviously, in order to describe a bound
state we seek for normalizable solutions
\begin{eqnarray}
\int_0^\infty \left[ u(r)^2+w(r)^2\right]=1, 
\end{eqnarray}
which impose conditions on the wave functions both at infinity as well as
at the origin.

The set of equations~(\ref{ref:CE}) must be accompanied by asymptotic conditions
at infinity. As in previous sections, we have a linear potential at large
distances for both channels and therefore the wave function is a linear
combination of regular and irregular Airy functions. Once we have discard the
irregular function at long  distances, Bi(z), the wave functions at
infinity have the following behaviour
\begin{eqnarray}
u(r) &\rightarrow& A_{S}\, \mbox{Ai}(z), \nonumber \\
w(r) &\rightarrow& A_{D}\, \mbox{Ai}(z),
\label{ref:ABC}
\end{eqnarray}
with $z$ defined in Eq.~(\ref{eq:z_def}) and $A_{S}$ is the normalization
factor and the asymptotic D/S ratio parameter is defined by
$\eta=A_{D}/A_{S}$.  Ideally, one would integrate the Schr\"odinger
equation taking its solutions at infinity, Eq.~(\ref{ref:ABC}), which
depend on the bound energy and $\eta$. The singular structure of the
problem at short distances requires a specific analysis of the coupled
equations as it has been done extensively
elsewhere~\cite{Cordon:2009pj} and we adapt in 
Appendix~\ref{sec:shortdistances} for our particular case. The result
amounts to integrate from infinity for the physical value of
$M_{J/\psi}$ and $\eta$. Generally, the solutions diverge strongly at
the origin, so that the normalization of the state is
precluded. However, there is a particular value of $\eta$ which
guarantees that the wave function becomes normalizable.  Then, if one
imposes the regularity condition at the origin one will determine
$\eta$ and therefore the wave function of the bound state. In
practice, however, the converging solution is rather elusive since
integrated-in solutions quickly run into diverging solution due to the
round-off errors and dominate over the converging solution.

According to Ref.~\cite{Cordon:2009pj} one may proceed as follows.
One can impose different auxiliary short distance boundary conditions
corresponding to a choice of regular solutions at the origin
\begin{eqnarray}
u(r_{c})=0 & \mbox{(BC1)}, \nonumber \\
u'(r_{c})=0 & \mbox{(BC2)}, \nonumber \\
w(r_{c})=0 & \mbox{(BC3)}, \nonumber \\
w'(r_{c})=0 & \mbox{(BC4)}, \nonumber \\
u(r_{c})-\sqrt{2}w(r_{c})=0 & \mbox{(BC5)}, \nonumber \\
u'(r_{c})-\sqrt{2}w'(r_{c})=0 & \mbox{(BC6)}.
\label{ref:BC}
\end{eqnarray}

All these boundary conditions must predict the same value of $\eta$ at
some value of the cut-off radius, $r_{c}$, the precise convergence
value corresponds to the particular choice.  As in other
Refs.~\cite{PavonValderrama:2005gu} we find convergence for the
boundary conditions BC5 and BC6 for larger cut-off radii, which
improves the numerical results.

To calculate the D/S asymptotic ratio, $\eta$, it is convenient to use
the superposition principle of boundary
conditions~\cite{PavonValderrama:2005gu} to write
\begin{equation}
\begin{split}
& u(r)=u_{S}(r)+\eta u_{D}(r), \\
& w(r)=w_{S}(r)+\eta w_{D}(r), 
\end{split}
\end{equation}
where $(u_{S},w_{S})$ and $(u_{D},w_{D})$ correspond to the boundary
conditions at infinity, Eq.~(\ref{ref:ABC}) with $A_{S}=1$ and
$A_{D}=0$ and with $A_{S}=0$ and $A_{D}=1$, respectively. Through that
decomposition the boundary conditions, BC1-BC6, can be rewritten as
algebraic expressions for $\eta$. For instance, if we use the BC6
boundary condition we get 
\begin{equation}
\eta=-\frac{u'_{S}(r_{c})-\sqrt{2}\,w'_{S}(r_{c})}{u'_{D}(r_{c})-\sqrt{2}\,w'_{D
} (r_ {c})}.
\end{equation}
Once $\eta$ has been calculated, the wave function of the bound state
is completely determined by the normalization factor $A_{S}$
\begin{eqnarray}
u(r) &=& A_{S}(u_{S}+\eta\,u_{D}), \nonumber \\ 
w(r) &=& A_{S}(w_{S}+\eta\,w_{D}),
\end{eqnarray}
in which $A_{S}$ is obtained normalizing the wave function to one
\begin{equation}
A_{S}^{2}\int_{0}^{\infty}[(u_{S}+\eta\,u_{D})^{2}+(w_{S}+\eta\,w_{D})^{2}]dr=1.
\end{equation}

The above procedure can be undertaken for the ground state of the
system if its energy is known. Now, if we want to calculate the
excited states of the system we must impose the orthogonality
condition between wave functions of states with different energy
together with the regularity condition at the origin.

For a regular potential the orthogonality between wave functions of
states with different energy is a property of the solutions of the
Schr\"odinger equation and it always holds. However, if the potential
contains some attractive singular contribution we have to impose
explicitly the orthogonality between wave functions.

Thus, given the ground state and one excited state, the orthogonality
condition can be written as
\begin{equation}
\int_{0}^{\infty}dr \left[u_{0}(r)u_{m}(r)+w_{0}(r)w_{m}(r)\right]=0,
\end{equation}
where it is useful to rewrite the above expression through a Lagrange identity
\begin{equation}
\left[u'_{0}u_{m}-u_{0}u'_{m}+w'_{0}w_{m}-w_{0}w'_{m}\right]\Big|_{0}^{
\infty}=0.
\label{ref:OC}
\end{equation}
Note that any individual term in the integrand is actually divergent,
because of the dominance of the singular solutions at the origin.  At
very short distances, the orthogonality between wave functions and the
regularity condition of them have been imposed at a certain cut-off
radius, $r_c$. Of course, we always check that the numerical
calculation is stable against suitable changes of the short distance
cut-off so that the range $r_c \sim 0.01 - 0.3 {\rm fm}$ is
sufficient. In that situation, the orthogonality condition,
Eq.~(\ref{ref:OC}), can be written as
\begin{equation}
\begin{split}
&u'_{0}(r_{c})u_{m}(r_{c})+w'_{0}(r_{c})w_{m}(r_{c}) \\ 
&
=u_{0}(r_{c})u'_{m}(r_{c})+w_{0}(r_{c})w'_{m}(r_{c}),
\end{split}
\end{equation}
and combining this expression with the corresponding one of the
boundary conditions, Eq.~(\ref{ref:BC}), we obtain in the case of the
boundary condition BC6 
\begin{equation}
\frac{w'_{m}(r_{c})}{\sqrt{2}\,u_{m}(r_{c})+w_{m}(r_{c})}=\frac{w'_{0}(r_{c})}{
\sqrt{2}\,u_{0}(r_{c})+w_{0}(r_{c})}.
\end{equation}
and similarly for all other auxiliary boundary conditions.  Obviously
in this case the $D/S$ mixing of the excited state is determined from
the requirement of regularity at the origin,
\begin{equation}
\eta_m=-\frac{u'_{S,m}(r_{c})-\sqrt{2}\,w'_{S,m}(r_{c})}{u'_{D,m}(r_{c})-\sqrt{2
}\,w'_{D,m}(r_{c})}.
\end{equation}

\begin{figure*}[ttt]
\begin{center}
\epsfig{figure=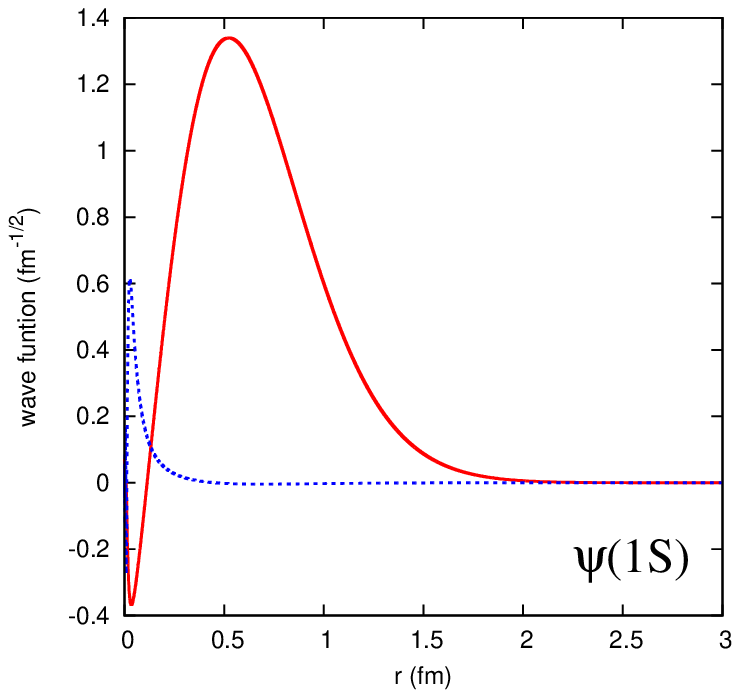,height=4cm,width=4cm}
\epsfig{figure=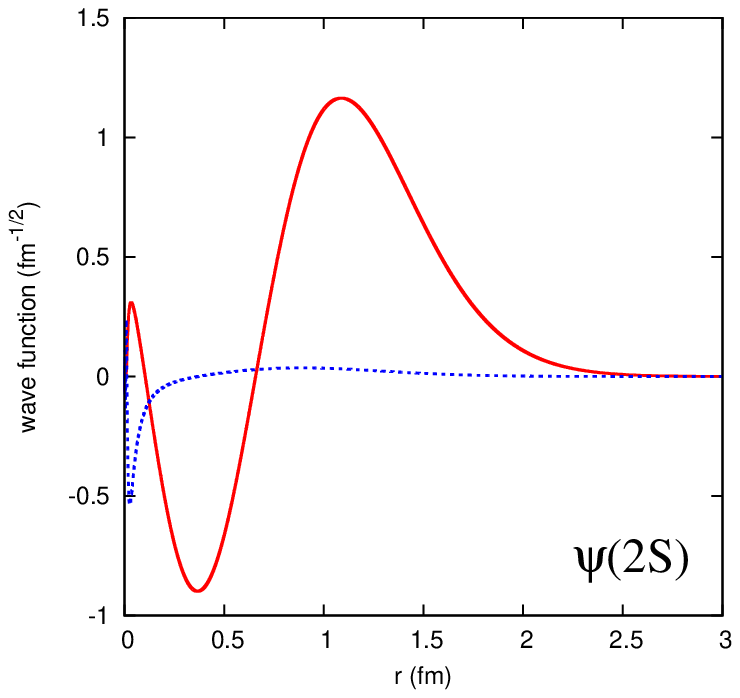,height=4cm,width=4cm}
\epsfig{figure=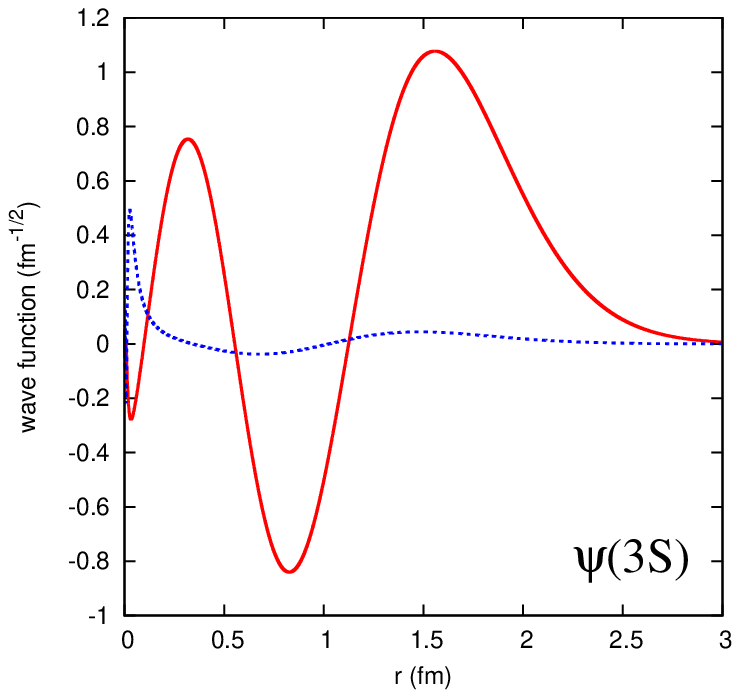,height=4cm,width=4cm}
\epsfig{figure=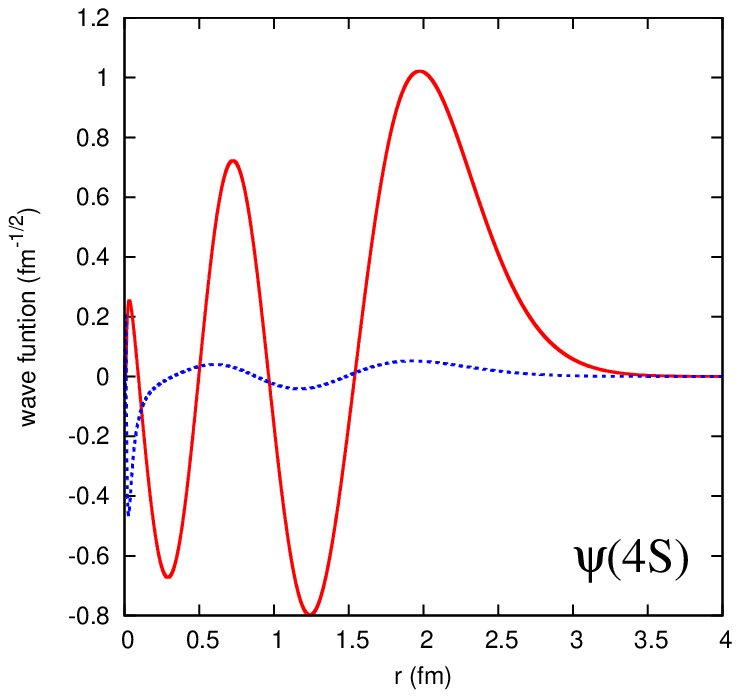,height=4cm,width=4cm} \\
\epsfig{figure=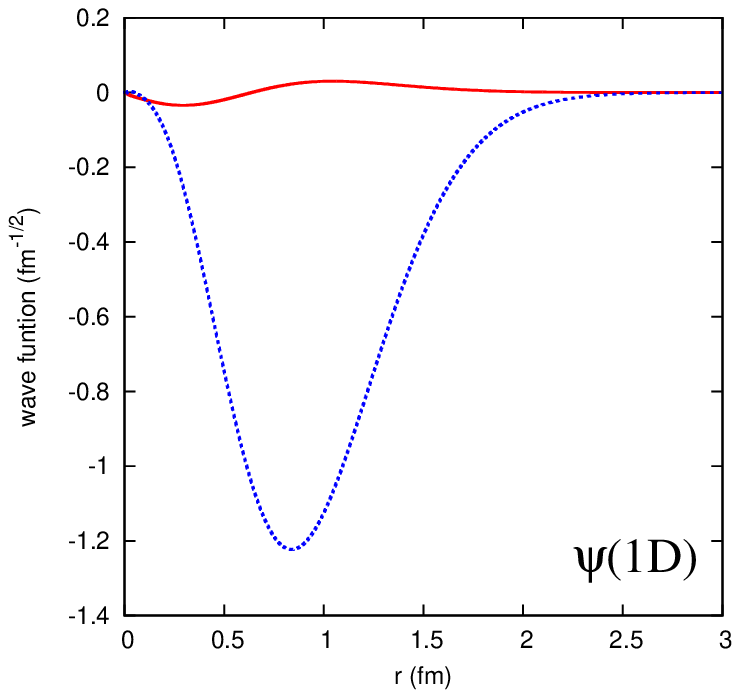,height=4cm,width=4cm}
\epsfig{figure=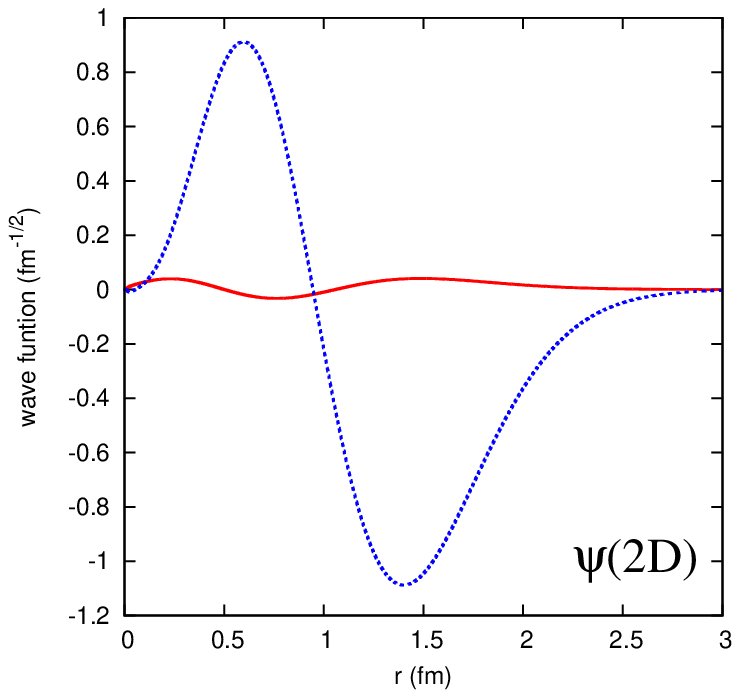,height=4cm,width=4cm}
\epsfig{figure=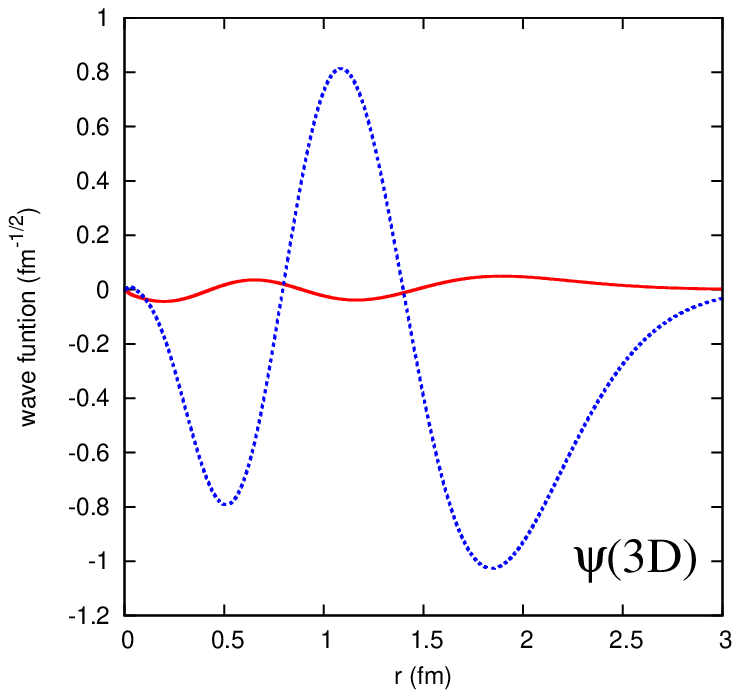,height=4cm,width=4cm}
\epsfig{figure=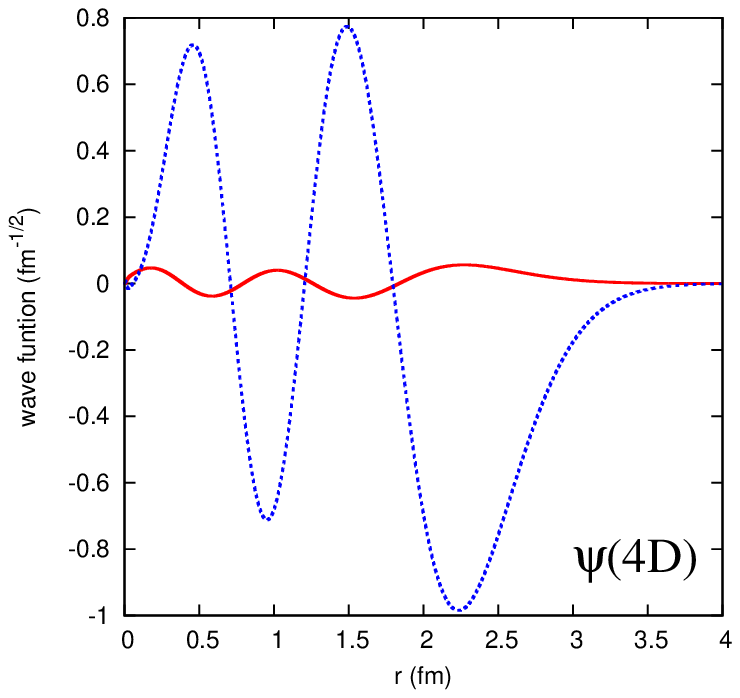,height=4cm,width=4cm}
\caption{\label{fig:reswfTM} S and D-wave functions of the different states of
charmonium calculated through the coupled toy model.}
\end{center}
\end{figure*}

We will also compute other observables. The results are presented in
Table~\ref{tab:toy-coupled}. In the numerical integration
a certain maximum radious has to be selected to imposed the asymptotic
boundary conditions. Due to the long range coulomb force the values
of the asymptotic parameters are slowly convergent and we quote the
values for $r_{\rm max}=20\,{\rm fm}$.
The corresponding wave functions can be
looked up in Fig.~\ref{fig:reswfTM}.

As we see the description of the spectrum is not particularly
accurate, but one should take into account that we are only providing
the $J/\psi$ mass as an input, as well as $\alpha_s$, $m_c$ and
$\sigma$, for which we take generally accepted values. The present
analysis suggests that exploiting these ideas one may find a reduction
of parameters in quark models. As we will see in later Sections,
rather than being a mathematical curiosity, these correlations which
are unveiled by the renormalization approach actually are embodied in
quark models in a much less transparent way. 

\begin{table*}[t!]
\begin{center}
\begin{tabular}{lccccccr}
\hline
\hline
State & M (MeV) & $\eta$ & $A_{s}$ $(fm^{-1/2})$ & & $\left\langle\right.\!\!
r^{2}\!\! \left.\right\rangle^{1/2}$ $(fm)$ & & 
$P_{D}$ $(\%)$ \\[2ex]
$J/\psi$ & $3096.916^{\dagger}$ & $-0.0037$ & $4.01$ & & $0.637$ & & 
$1.52$ \\
$\psi(1D)$ & $3577.7$ & $-19,71$ & $0.084$ & & $0.93$ & & $99.91$
\\
$\psi(2S)$ & $3634.2$ & $+0.015$ & $3.05$ & & $1.05$ & & $1.24$ \\
$\psi(2D)$ & $3995.9$ & $-17.73$ & $0.104$ & & $1.29$ & & $99.84$
\\
$\psi(3S)$ & $4065.1$ & $+0.028$ & $2.66$ & & $1.40$ & & $1.15$ \\
$\psi(3D)$ & $4368.7$ & $-15.70$ & $0.12$ & & $1.60$ & & $99.75$
\\
$\psi(4S)$ & $4443.1$ & $+0.039$ & $2.44$ & & $1.70$ & & $1.12$ \\
$\psi(4D)$ & $4710.8$ & $-14.09$ & $0.13$ & & $1.88$ & & $99.65$
\\
$\psi(5S)$ & $4787.4$ & $+0.048$ & $2.29$ & & $1.98$ & & $1.14$ \\
\hline
\hline
\end{tabular}
\caption{\label{tab:toy-coupled} Different observables computed through the
coupled toy model and applied to charmonium spectrum.}
\end{center}
\end{table*}

\section{Charmonium with form factors}
\label{sec:oldmodel}

In the previous sections we have presented how one may incorporate a
desirable short distance insensitivity into several simplified models
for quarkonium. In this section we give a brief description of the
specific charmonium model which fits a wider and successful
phenomenology (see Ref.~\cite{Vijande:2005eu} for further details)
paying special attention to the form factors introduced to regulate
the unpleasent short distance singularities.  Our intention is to
re-analyse the model by trading these form factors into a less model
dependent set of renormalization conditions. While the model furnishes
unequal quark species we will restrict for illustration purposes to
the case of charmonium. 


The corresponding potentials for the $\bar{q}q$ system stem from the
non-relativistic reduction of OGE and the confinement component. We
separate the central, tensor and spin-orbit pieces, as follows: 
\begin{itemize}
\item One-Gluon Exchange
\begin{equation}
\begin{split}
 &
V_{OGE}^{C}(\vec{r})=\frac{1}{4}\alpha_{s}
(\vec{\lambda}_{1}^{c}\cdot\vec{\lambda}_{2}^{c})
 \left[ \frac{1}{r}-\frac{1}{6m^2} 
(\vec{\sigma}_{1}\cdot\vec{\sigma}_{2}) 
\frac{e^{-r/r_{0}(\mu)}}{rr_{0}^{2}(\mu)}\right], \\
 & 
V_{OGE}^{T}(\vec{r})=-\frac{1}{16}\frac{\alpha_{s}}{m^2}(
\vec{\lambda}_{1}^{c}\cdot\vec{\lambda}_{2}^{c}) \\ 
& \times \left[ 
\frac{1}{r^{3}}-\frac{e^{-r/r_{g}(\mu)}}{r}\left( 
\frac{1}{r^{2}}+\frac{1}{3r_{g}^{2}(\mu)}+\frac{1}{rr_{g}(\mu)}\right)
\right]S_{12}, \\
 &
\begin{split}
 V_{OGE}^{SO}(\vec{r})= &  
-\frac{3}{8}\frac{\alpha_{s}}{m^2}(
\vec{\lambda}_{1}^{c}\cdot\vec{\lambda}_{2}^{c}) \\ 
& \times
\left[\frac{1}{r^{3}}-
\frac{e^{-r/r_{g}(\mu)}}{r^{3}}\left(1+\frac{r}{r_{g}(\mu)}\right)\right] \vec{L} \cdot \vec{S}.
\end{split}
\end{split}
\end{equation}
\item Confinement
\begin{equation}
\begin{split}
 &
 V_{CON}^{C}(\vec{r})=\left[ -a_{c}(1-e^{-\mu_{c}r})+\Delta \right] 
(\vec{\lambda}_{1}^{c}\cdot\vec{\lambda}_{2}^{c}), \\
 &
 \begin{split}
 V_{CON}^{SO}(\vec{r})= &
-(\vec{\lambda}_{1}^{c}\cdot\vec{\lambda}_{2}^{c}) 
\frac{a_{c}\mu_{c}e^{-\mu_{c}r}}{4m^2 r}(6 -8 a_{s}) \vec{L} \cdot \vec{S} ,
 \end{split}
 \end{split}
\label{eq:Conf_V}
 \end{equation}
\end{itemize}
where $r_{0}(\mu)=\hat{r}_{0}\frac{\mu_{nn}}{\mu}$ and
$r_{g}(\mu)=\hat{r}_{g}\frac{\mu_{nn}}{\mu}$ are short distance regulators
and $\mu$ stand for the reduced quark mass. The contact term of the
central potential of  one gluon exchange is regularized in a suitable way by
replacing the Dirac delta function by a Yukawa form  
\begin{eqnarray}
\delta (\vec r ) \to \frac{1}{4\pi r_0^2} \frac{e^{-r/r_0}}{r}.
\end{eqnarray}
In table~\ref{tab:parameters} we show
the model parameters constrained by the light quark phenomenology and also
appear in the potentials operating in the heavy quark sector.

Regarding the confinement interaction~\footnote{Our expressions imply
  the validity of Casimir scaling both for the perturbative (tree
  level) as well as the non-perturbative (confinement) contributions
  to the potential. A recent perturbative calculation to three loops
  displays violations to this Casimir scaling~\cite{Anzai:2010td},
  although nothing is implied for the non-perturbative contribution.},
$\Delta $ is a global constant adjusted to fit the origin of
energies, $a_{c}$ and $\mu_{c}$ are model parameters. At short distances
this potential presents a linear behavior with an effective confinement strength
$\sigma=-a_{c}\,\mu_{c}\,(\vec{\lambda}^{c}_{1} \cdot \vec{\lambda}^{c}_{2})$
while it becomes constant at large distances. This type of potential shows a
threshold defined by
\begin{equation}
V_{thr}=\{-a_{c}+ \Delta\}(\vec{\lambda}^{c}_{1}\cdot \vec{\lambda}^{c}_{2}).
\label{eq:thres-conf}
\end{equation}

\begin{table}[t!]
\begin{center}
\begin{tabular}{c|cc}
 \hline
 \hline
 Quark mass & $m_{c}$ (MeV) & $1763$ \\
 \hline
 Confinement & $a_{c}$ (MeV) & $507.4$ \\
	       & $\mu_{c}$ $(\mbox{fm}^{-1})$ & $0.576$ \\
	       & $\Delta$ (MeV) & $184.432$ \\
	       & $a_{s}$ & $0.81$ \\
 \hline
 OGE & $\alpha_{0}$ & $2.118$ \\
     & $\Lambda_{0}$ $(\mbox{fm}^{-1})$ & $0.113$ \\
     & $\mu_{0}$ (MeV) & $36.976$ \\
     & $\hat{r}_{0}$ (fm) & $0.181$ \\
     & $\hat{r}_{g}$ (fm) & $0.259$ \\
 \hline
 \hline
\end{tabular}
\caption{\label{tab:parameters} Quark model parameters.}
\end{center}
\end{table}

By construction, no $\bar{q}q$ bound states can be found for energies
higher than this threshold. Actually, the potential suffers a transition from a
color string configuration between two static color sources into a pair of
static mesons due to the breaking of the color flux-tube and the most
favoured subsequent decay into hadrons. In a more general set up, a dynamical
coupling to the formed meson pair should be included. 

The wide energy range covered by a consistent description of light, strange and
heavy mesons requires an effective scale-dependent strong coupling constant that
cannot be obtained from the usual one-loop expression of the running coupling
constant which diverges at $Q^{2}=\Lambda_{QCD}^{2}$. Following previous
work~\cite{Vijande:2005eu}, we use a frozen coupling constant such as 
\begin{equation}
\alpha_{s}(\mu)=\frac{\alpha_{0}}{\ln\left( 
\frac{\mu^{2}+\mu_{0}^{2}}{\Lambda_{0}^{2}} \right)},
\end{equation}
where $\mu$ is the reduced mass of the $\bar{q}q$ pair and $\alpha_{0}$,
$\mu_{0}$ and $\Lambda_{0}$ are parameters of the model determined by a global
fit to the hyperfine splitting from the light to the heavy quark sector. In our
case this yields a value of $\alpha_s=0.288$. 

We note that the regulators in the tensor and spin-orbit terms of the one gluon
exchange potential as well as the smeared $\delta$-function of the central
component are introduced just because tensor and spin-orbit terms become
singular at short distances. As a consequence the collapse of the system which
would occur for the pure unregularized $\delta$-function is prevented. The
previous model describes successfully the meson
spectroscopy~\cite{Vijande:2005eu} and provides a good agreement with the
experimental data of different physical observables as the strong and radiative
decays of charmonium performed in Ref.~\cite{Segovia:2008zz}.

As we have discussed above it is interesting to inquire the real need
of form factors which, while providing an acceptable phenomenology,
are introduced {\it ad hoc} to avoid the short distance
divergences. Indeed, if the short distance regulators are removed,
i.e. $r_0,r_g \to 0$, the expressions of our potentials become
Eq.~(\ref{eq:OGE_unreg}) for the OGE piece and the confinement part
remains the same as in Eq.~(\ref{eq:Conf_V}) as this latter
contribution does not depend on regulators. We will use these
expressions for the potentials in the renormalization scheme. Note
that we have also discarded the Dirac delta functions. While this may
seem weird, these are distributions which are not seen in the compact
support test functions implicitly implied by the boundary
condition regularization. This also applies to any derivatives of the
Dirac delta function. This result was
suggested~\cite{PavonValderrama:2005wv} and explicitly checked by
using a momentum space regularization with so-called
counterterms~\cite{Entem:2007jg}.

Moreover, the form factors are naturally adjusted in order to reproduce some
well established physical observables, such as the $J/\psi$ mass and its
leptonic width. What will be shown below is that one can actually treat 
these observables as suitable renormalization conditions, without any specific
need of form factor regulators. 

\section{Renormalized charmonium model}
\label{sec:renormodel}

\subsection{Masses (uncoupled case)}

\begin{figure}[ttt]
\begin{center}
\epsfig{figure=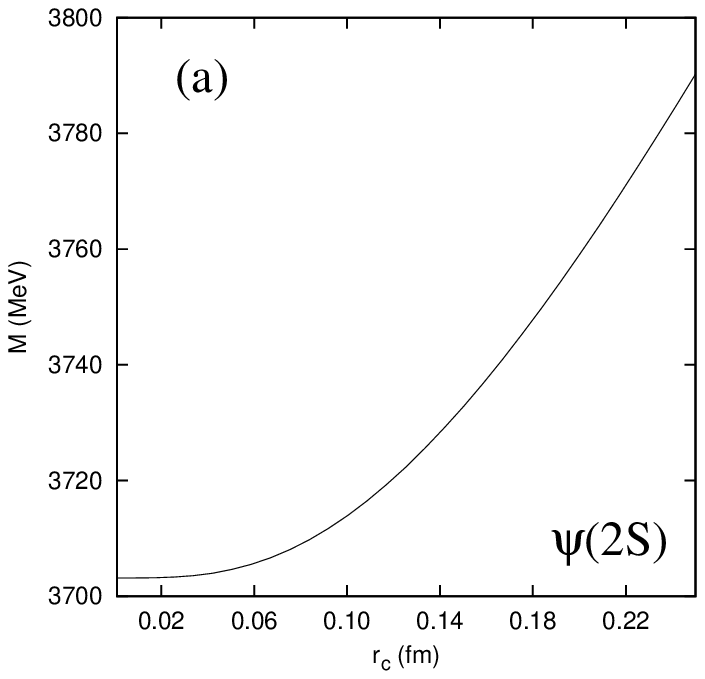,height=6.5cm,width=7cm}
\epsfig{figure=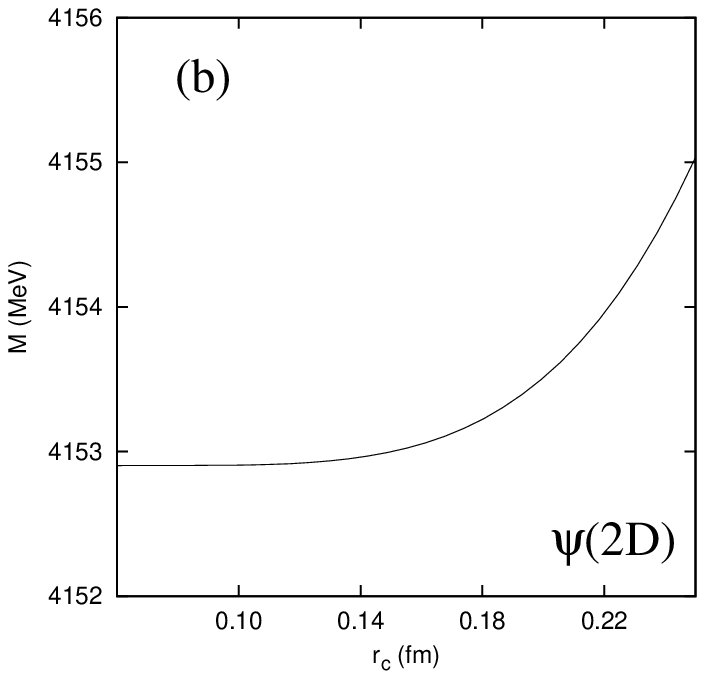,height=6.5cm,width=7cm}
\end{center}
\caption{Dependence of the mass (in MeV) $2^3S_1$ (upper panel) and
  $2^3D_1$ (lower panel) excited charmonium states on the short
  distance cut-off, $r_c$ (in fm).}
\label{fig:cut-off-masses}
\end{figure}

In this section we carry out the renormalization, presented previously
along the lines described in detail in Section~\ref{sec:toy}, for the
unregularized model potential.  While the tensor force induces a
mixing between S and D-waves we will in a first step neglect such a
mixing. This implies that both ground states are completely unrelated
and renormalization is pusued independently.  As expected, the
charmonium masses in this scheme depend on the short distance cut-off,
$r_{c}$.  Fig.~\ref{fig:cut-off-masses} shows this dependence for the
first radial excitations of $S$ and $D$-wave charmonium states. One
can see that at some value of $r_{c}$ the masses do not depend on the
short distance cut-off.

In Table~\ref{tab:masses} the masses predicted by renormalized
charmonium model and the standard constituent quark model (with form
factors) are displayed. We find a perfect agreement between both
schemes. This provides confidence on the way the original model took
into account the unknown short distance dynamics, on the one hand, and
also on the irrelevance of those form factors for excited states as
long as the ground state mass is fixed.

\begin{table}[t!]
\begin{center}
\begin{tabular}{cccccc}
\hline
\hline
State & n & $M_{RSC}$ (MeV) & $M_{CQM}$ (MeV) & $M_{exp}$ (MeV)
& Ref. \\
\hline
$^{3}S_{1}$ 	& $1$ & $3096^{\dagger}$ & $3096$ & $3096.916\pm0.011$ &
\cite{Nakamura:2010zzi} \\
		& $2$ & $3703$ & $3703$ & $3686.093\pm0.034$ &
\cite{Nakamura:2010zzi} \\
		& $3$ & $4097$ & $4097$ & $4039.6\pm4.3$ &
\cite{Nakamura:2010zzi} \\
		& $4$ & $4389$ & $4389$ & $4361\pm9\pm9$
& \cite{PhysRevLett.99.142002} \\
		& $5$ & $4614$ & $4614$ & $4634^{+8+5}_{-7-8}$ &
\cite{PhysRevLett.101.172001} \\[2ex]
$^{3}D_{1}$	& $1$ & $3796^{\dagger}$ & $3796$ & $3772.92\pm0.35$ &
\cite{Nakamura:2010zzi} \\ 
		& $2$ & $4153$ & $4153$ & $4153\pm3$ &
\cite{Nakamura:2010zzi} \\ 
		& $3$ & $4426$ & $4426$ & $4421\pm4$ &
\cite{Nakamura:2010zzi} \\ 
		& $4$ & $4641$ & $4641$ & $4664\pm11\pm5$ &
\cite{PhysRevLett.99.142002} \\
\hline
\hline
\end{tabular}
\end{center}
\caption{\label{tab:masses} Predicted charmonium masses (in MeV) within the
renormalization scheme (RSC) and potential model with form factors (CQM).}
\end{table}

As it becomes clear from the expressions of the potential, the perturbative and
non-perturbative contributions dominate at short and large distances
respectively. Therefore, we want to study now the dependence of the mass with
respect to two important model parameters, the strong coupling constant,
$\alpha_{s}$, and our effective string tension of the confinement potential,
$\sigma$, featuring those short and long distance effects.

\begin{figure}[ttt]
\begin{center}
\epsfig{figure=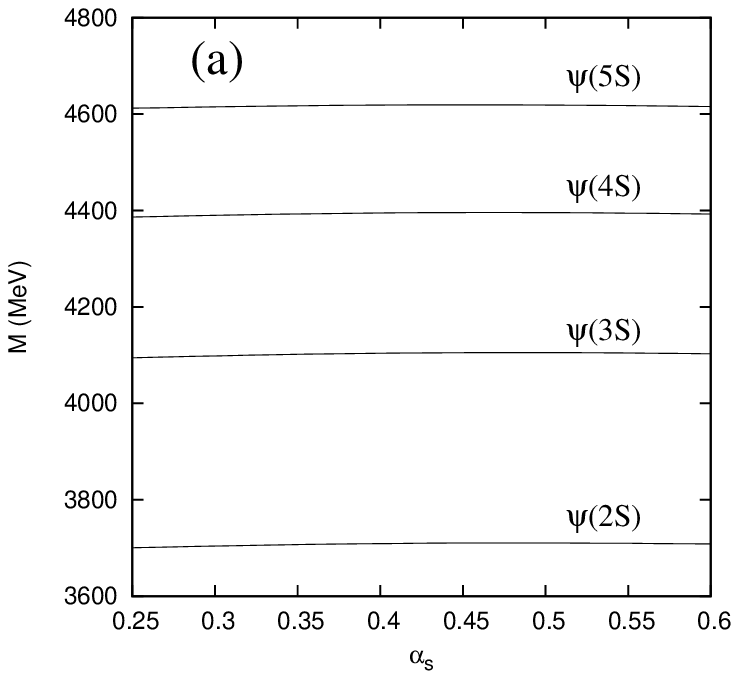,height=6.5cm,width=8cm}
\epsfig{figure=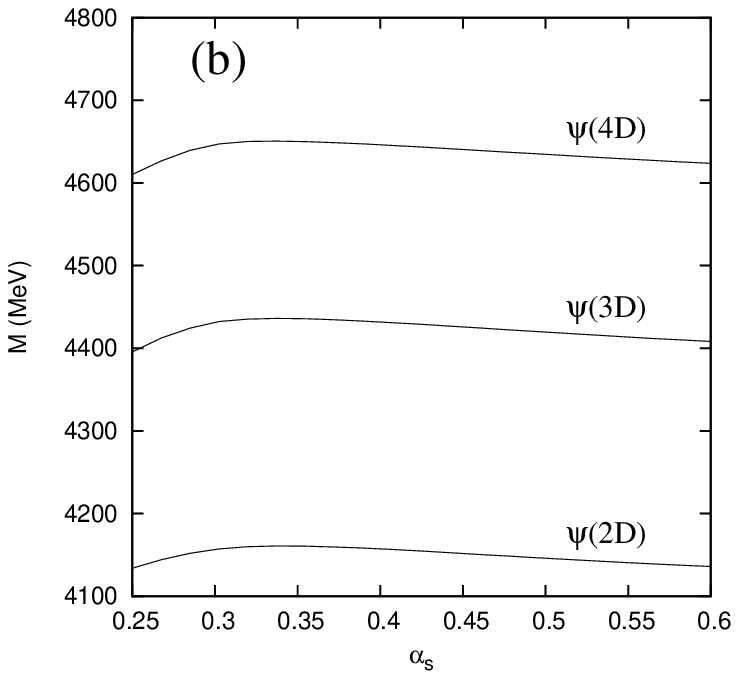,height=6.5cm,width=8cm}
\end{center}
\caption{Mass (in MeV) dependence on the strong coupling constant of excited
charmonium states.}
\label{fig:masses-vs-als}
\end{figure}

The dependence on $\alpha_s$ while fixing the threshold for confinement, $V_{\rm
thr}$ in Eq.~(\ref{eq:thres-conf}), can be seen in Fig.~\ref{fig:masses-vs-als}
for $S$ and $D$-wave states. For $S$-wave states the mass changes about $15 \,{\rm
MeV}$ meaning a $0.4\%$ in the total mass and we see a flattening
behaviour along the considered range of $\alpha_{s}$. We find a similar trend
for the $D$-wave states, although in this case the change in mass is larger than
in the case of $S$-wave states, the picture shows again a rather clear plateau.

\begin{figure}[ttt]
\begin{center}
\epsfig{figure=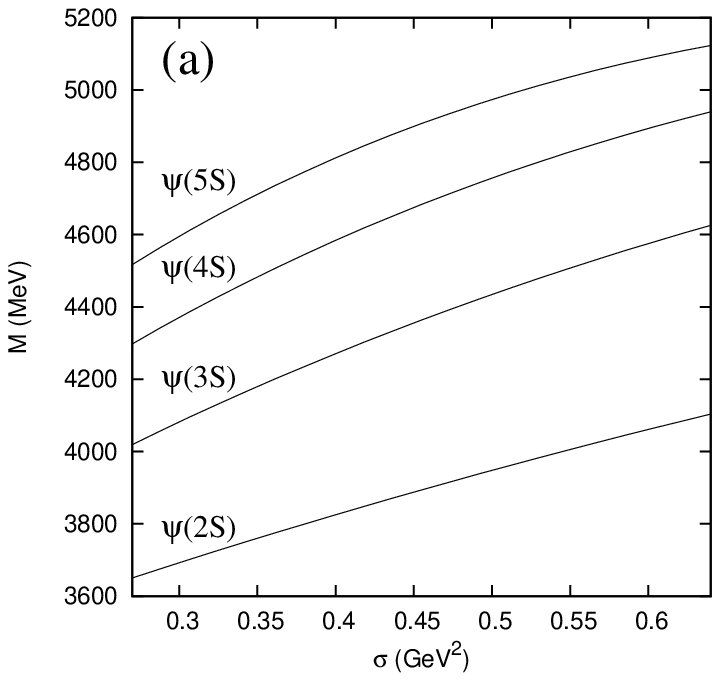,height=6.5cm,width=8cm}
\epsfig{figure=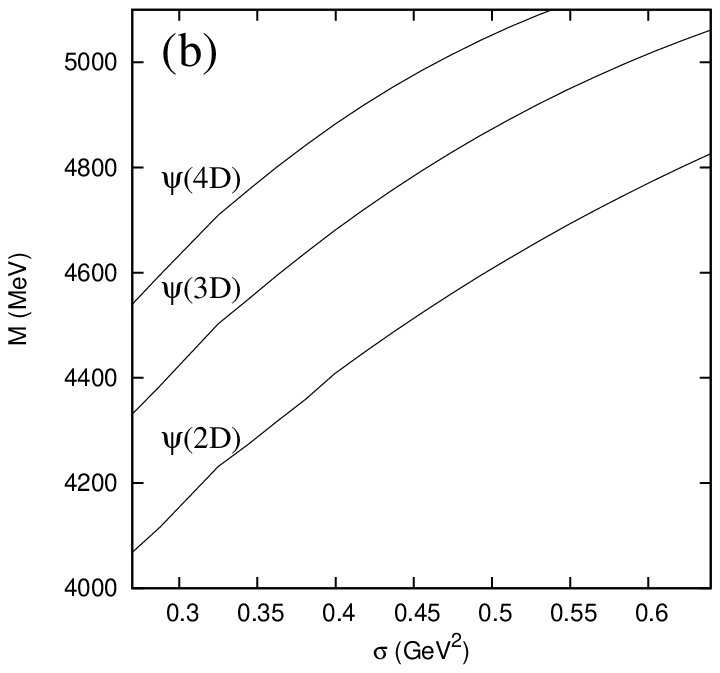,height=6.5cm,width=8cm}
\end{center}
\caption{Dependence of the mass (in MeV) excited charmonium states on the
effective string tension of our confinement potential, $\sigma=-a_{c}\,\mu 
_{c}\,(\vec{\lambda}^{c}_{i}\cdot \vec{\lambda}^{c}_{j})$ (in ${\rm GeV}^{2}$).}
\label{fig:masses-vs-sigma}
\end{figure}

We now turn to the mass dependence on the effective string tension of
our confinement potential. Again and for clarity of presentation we
fix the threshold for confinement, $V_{\rm thr}$ in
Eq.~(\ref{eq:thres-conf}).  Fig.~\ref{fig:masses-vs-sigma} shows such
a dependence for the $S$ and $D$-wave states. The range for the
effective string tension is in percentage level equal than the range
of the strong coupling constant and we can see that the masses change
on the hundreds of MeV. We can conclude that at least the masses of
excited states are dominated by the confinement potential as long as
the ground state mass is kept to its physical value. In
Appendix~\ref{sec:fitting} we further analyze the sensitivity of the
renormalized model parameters.

\subsection{Leptonic widths}

We now focus on the leptonic widths. This will illustrate the interesting
subject of the wave function renormalization. A complete calculation of
$V\mbox{(vector meson)}\rightarrow e^{+}e^{-}$ widths involves radiative and
relativistic contributions. In the heavy quark limit one can write the leptonic
decay for $S$-wave states
as~\cite{Barbieri:1975eu}
\begin{equation}
\Gamma\left(n^{3}S_{1} \rightarrow
e^{+}e^{-}\right)=\frac{4\alpha^{2}e^{2}_{c}|R_{nS}(0)|^2}{M^{2}_{n}}\left(1-
\frac{16\alpha_s}{3\pi}\right),
\end{equation}
where $e_{c}=2/3$ and $M_{n}$ is the mass of the charmonium state. Decay widths 
depend on the value of the wave function at short distances. Of course while 
the decay may be triggered by a short distance operator, we may predict decay 
{\it ratios} as 
\begin{equation}
{\cal R}= \frac{\Gamma\left(n^{3}S_{1}\rightarrow 
e^{+}e^{-}\right)}{\Gamma\left(1^{3}S_{1}\rightarrow e^{+}e^{-}\right)}= 
\frac{|R_{nS}(0)|^2}{|R_{1S}(0)|^2}\frac{M^{2}_{1}}{M^{2}_{n}}.
\end{equation}
This corresponds in practice to implement a common wave function renormalization
which factors out in the ratio. 

\begin{figure}[ttt]
\begin{center}
\epsfig{figure=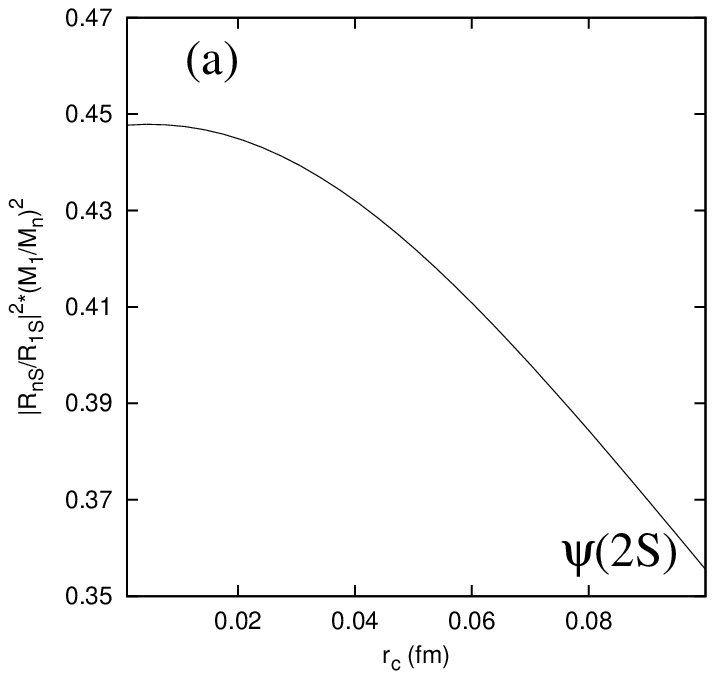,height=6.5cm,width=8cm}
\epsfig{figure=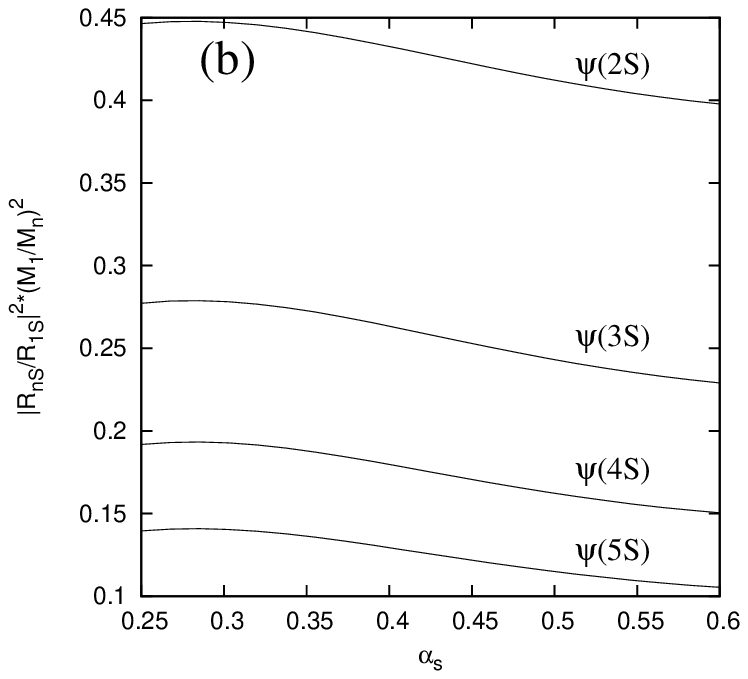,height=6.5cm,width=8cm}
\end{center}
\caption{Top picture shows the dependence of the ratio ${\cal R}$ with respect 
to short distance cut-off, $r_{c}$. Bottom picture shows the dependence of the
ratio ${\cal R}$ for the excited S-wave charmonium states on the strong
coupling constant, $\alpha_{s}$.}
\label{fig:lepwidth-vs-als}
\end{figure}

Fig.~\ref{fig:lepwidth-vs-als}(a) shows the dependence of ${\cal R}$ with
respect to the short distance cut-off for the first radial excitation of
$S$-wave states. At some range of $r_{c}$, the ratio does not depend on its
value.  Fig.~\ref{fig:lepwidth-vs-als}(b) shows the dependence of the ratio for
the $S$-wave states along the range of strong coupling constant. This range is
the same as in the case of studying masses. We find a stronger dependence on the
strong coupling constant as expected since the leptonic decay is a short range
observable.

\subsection{Inclusion of coupled channels}

Finally, we take into account the role played by the tensor force which, as
we know from previous sections, is the responsible of the mixing between S and
D-waves in the case of charmonium. The interesting aspect of applying
renormalization procedure to S-D coupled case is that using just one
renormalization condition, the mass of the $J/\psi$, we can predict all S and
D-wave mesons.

The procedure is the same than in Sec.~\ref{sec:toy-coupled} with only
one difference, where our confining interaction is given by
Eq.~(\ref{eq:Conf_V}), which flattens out at large distances. So in
this case  the asymptotic boundary conditions become 
\begin{eqnarray}
u(r) &\rightarrow& A_{S}\,e^{-\kappa r}, \nonumber \\
w(r) &\rightarrow& A_{D}\,\left( 1+\frac{3}{\kappa r}+\frac{3}{(\kappa r)^2} \right)\,e^{-\kappa r},
\label{ref:ABC2}
\end{eqnarray}
with $\kappa^{2}= m_{c}(V_{thr}+2m_{c}-M)$ and $V_{thr}$ given by
Eq.~(\ref{eq:thres-conf}).

Results on the mass, the asymptotic D/S ratio parameter and D-wave
probability are presented in Table~\ref{tab:cqm-coupled1}. The
comparison between renormalization scheme and constituent quark model
with form factors is given in Table~\ref{tab:cqm-coupled2}. One can
see that the agreement is completely satisfactory. Essentially, this
proves that the form factors only provide the correct mass of
$J/\psi$. Once this is fixed the rest of the excited states with
either S- or D- wave character are predicted. This is the main result
of the present study.

\begin{table}[ttt]
\begin{center}
\begin{tabular}{ccccccr}
\hline
\hline
State & & M (MeV) & & $\eta$ & & $P_{D}$ $(\%)$ \\[2ex]
$J/\psi$ & & $3095.78^{\dagger}$ & & $-0.0115$ & & $0.146$
\\
$\psi(2S)$ & & $3703.19$ & & $+0.0105$ & & $0.221$ \\
$\psi(1D)$ & & $3795.93$ & & $-12.47$ & & $99.852$ \\
$\psi(3S)$ & & $4097.46$ & & $+0.0321$ & & $0.446$ \\
$\psi(2D)$ & & $4152.20$ & & $-9.49$ & & $99.605$ \\
$\psi(4S)$ & & $4389.10$ & & $+0.0522$ & & $0.758$ \\
$\psi(3D)$ & & $4425.43$ & & $-7.76$ & & $99.280$ \\
$\psi(5S)$ & & $4614.28$ & & $+0.0698$ & & $1.108$ \\
$\psi(4D)$ & & $4640.00$ & & $-6.70$ & & $98.925$ \\
\hline
\hline
\end{tabular}
\end{center}
\caption{\label{tab:cqm-coupled1} Masses (in MeV), the asymptotic D/S ratio
parameter and D-wave probabilities of charmonium states including S-D mixture.
The only input is $M_{J/\psi}$.}
\end{table}

\begin{table*}[t!]
\begin{center}
\begin{tabular}{rr|rrc|rrc|c}
\hline
\hline
Meson & n & \multicolumn{3}{|c|}{Renormalized scheme} & \multicolumn{3}{|c|}{Form factors scheme} & Exp. data \\
\hline
& & $\mathcal P_{^3\!S_{1}}\,(\%)$ & $\mathcal P_{^3\!D_{1}}\,(\%)$ & $M$ (MeV) & $\mathcal P_{^3\!S_{1}}\,(\%)$ & $\mathcal P_{^3\!D_{1}}\,(\%)$ & $M$ (MeV) & $M$ (MeV) \\[2ex]
$J/\psi$ 	& $1$ & $99.85$ & $0.15$ & $3096^{\dagger}$ & $99.96$ & $0.04$ & $3096$ & $3096.916\pm0.011$ \\
$\psi(2S)$	& $2$ & $99.78$ & $0.22$ & $3703$ & $99.96$ & $0.04$ & $3703$ & $3686.093\pm0.034$ \\
$\psi(3770)$	& $3$ & $0.15$ & $99.85$ & $3796$ & $0.03$ & $99.97$ & $3796$ & $3772.92\pm0.35$ \\ 
$\psi(4040)$	& $4$ & $99.55$ & $0.45$ & $4098$ & $99.94$ & $0.06$ & $4097$ & $4039.6\pm4.3$ \\
$\psi(4160)$	& $5$ & $0.39$ & $99.61$ & $4152$ & $0.06$ & $99.94$ & $4153$ & $4153\pm3$ \\ 
$X(4360)$	& $6$ & $99.24$ & $0.76$ & $4389$ & $99.91$ & $0.09$ & $4389$ & $4361\pm9\pm9$ \\
$\psi(4415)$	& $7$ & $0.72$ & $99.28$ & $4426$ & $0.09$ & $99.91$ & $4426$ & $4421\pm4$ \\ 
$X(4630)$	& $8$ & $98.89$ & $1.11$ & $4614$ & $99.88$ & $0.12$ & $4614$ & $4634^{+8+5}_{-7-8}$ \\
$X(4660)$	& $9$ & $1.08$ & $98.92$ & $4640$ & $0.11$ & $99.89$ & $4641$ & $4664\pm11\pm5$ \\
\hline
\hline
\end{tabular}
\end{center}
\caption{\label{tab:cqm-coupled2} Comparison of different properties of
charmonium between the renormalization scheme and the constituent quark model with form
factors, considering coupled channels in both cases.}
\end{table*}

\section{On the applicability of the renormalization procedure}
\label{sec:applic}

In most examples shown in this paper, we have always dealt with a
situation where it was possible to fix the energy of a bound state, so
that there is a mild dependence of excited states on the short
distance cut-off. This scaling analysis is carried out in
Appendix~\ref{sec:errors} providing an understanding of the features 
found numerically. 

Of course, there is the question when the present renormalization
procedure is guaranteed to work, i.e. can we always mock up our
ignorance at short distances by a given renormalization condition? On
the other hand it is also important to know if the answer is 
affirmative what does the renormalization tell about the true
solution?  

It is instructive to analyze first the case where renormalization can
be carried out at any level of approximation of the ``true''
potential.  We provide some insight into the renormalization problem
by analysing in some detail the bosonic string model
(BSM)~\cite{Alvarez:1981kc,Arvis:1983fp}. Such a model provides an
alternative approach to the physics of confinement from the point of
view of large distances (for a review see
e.g.~\cite{Lucha:1991vn}). It should be noted that with the constraint
$\alpha_s = \pi/16$, the Cornell potential corresponds to the large
distance expansion of the following potential
\begin{eqnarray}
V(r)= \sigma \sqrt{r^2 - r_0^2} = \sigma r - \frac{\sigma r_0^2}{2 r}
-\frac{\sigma r_0^4}{8 r^3} + \dots,
\label{eq:pot-bsm}
\end{eqnarray}
where $\sigma r_0^2 = \pi/6$. With the standard value for the string
tension $\sqrt{\sigma} = 0.420 \, {\rm GeV}$ one has $r_0 \sim 0.3 \,
{\rm fm}$. A feature of this potential is that for $r<r_0$ it becomes
purely imaginary. Moreover, lattice calculations~\cite{Necco:2001xg}
yield $4\alpha_s/3= 0.25(1)$, extremely close to the BSM choice of
$\alpha_s = \pi/16$ when the Cornell potential,
Eq.~(\ref{eq:pot-cornell}), is taken. It has been
found~\cite{Luscher:2002qv} that the transition from perturbative to
string behaviour is evident from the data and takes place at
surprisingly small distances~\footnote{This of course raises the
question on whether the perturbative OGE contribution,$-4
\alpha_s(\mu)/3 r$, should be added on top of the BSM-$1/r$ result or
it is actually dual to it, in which case a scale of $\mu=2 {\rm GeV}$
is implied, if one demands $\alpha_s(\mu) = \pi /16$ at LO in perturbation theory.}.

The BSM potential satisfies exact fundamental inequalities for the Wilson
loop exploiting reflection positivity~\cite{Bachas:1985xs,Nussinov:2000kn}, namely
\begin{eqnarray}
V'(r)>0 \qquad V''(r) \le 0.
\label{eq:concavity}
\end{eqnarray}
The latter identity is saturated by the linear potential. This is satisfied 
by  {\it any} inverse power series with  negative coefficients, i.e.
\begin{eqnarray}
V(r) = \sigma r + V_0 + \sum_{n=1}^\infty c_n /r^n,
\label{eq:ipower}
\end{eqnarray}
where one has $c_n \le 0$.  However, note that the converse need not
be true, namely Eq.~(\ref{eq:concavity}) does not necessarily imply
that {\it all} coefficients in Eq.~(\ref{eq:ipower}) are negative. 

\subsection{Renormalization of the bosonic string potential}
\label{sec:bsm}

In this section we deal with the renormalization of the bosonic-string potential
given by Eq.~(\ref{eq:pot-bsm}). Of course, this potential is questionable below
the critical radius where it becomes imaginary. On the other hand, if the large
distance expansion is truncated to a finite order we have to deal with
increasingly singular power divergences. We will illustrate how the
renormalization procedure works when the full potential is compared to the
series of singular potentials at the origin.

This example also shows how to treat singular potentials. Indeed, at
NNLO one encounters a $1/r^3$ short distance divergence. The problem
with such a singularity is that the energy becomes unbound from
below. To overcome this situation one can fix, as it was discussed in
Sec.~\ref{sec:toy}, the ground state energy and then, using the
orthogonality condition, the excited states can be predicted. Of
course, this procedure can also be carried out without a
singular potential, but just on the basis that the short distance
behaviour of the potential may not be fully reliable.

\begin{figure*}[ttt]
\begin{center}
\epsfig{figure=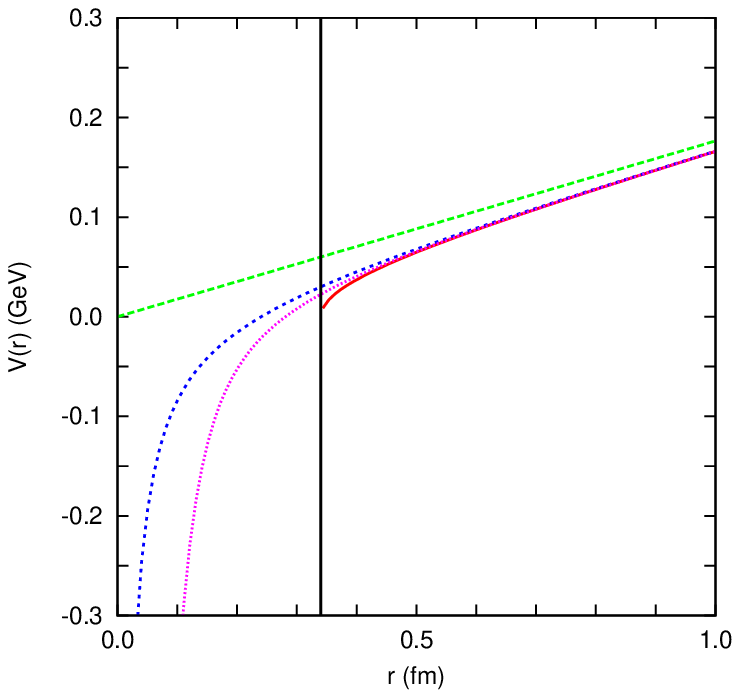,height=5cm,width=5.5cm}
\epsfig{figure=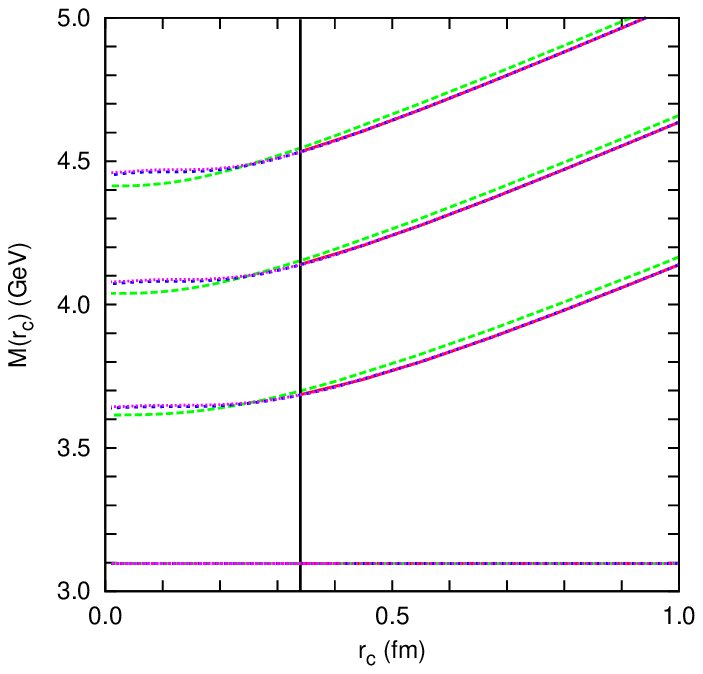,height=5cm,width=5.5cm}
\epsfig{figure=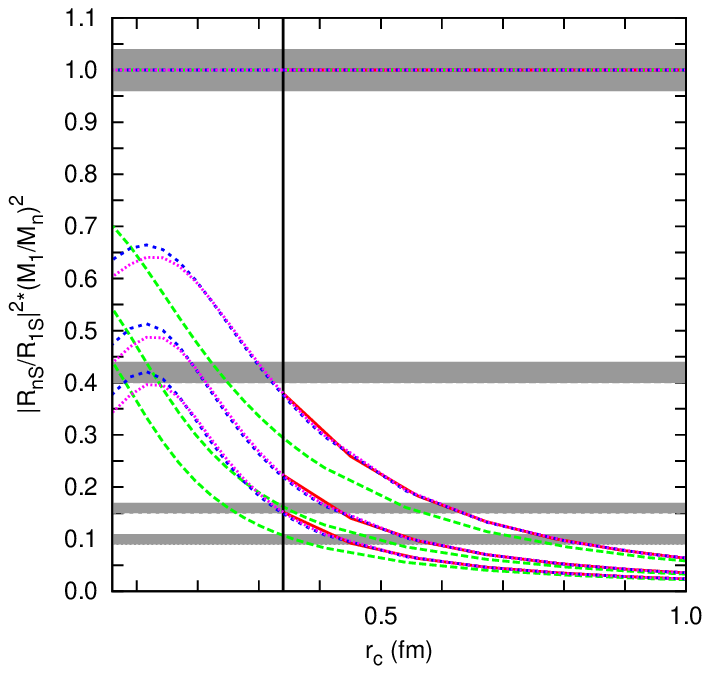,height=5cm,width=5.5cm}
\end{center}
\caption{{\bf Left panel:} The bosonic-string model potential (in GeV) and
  different terms contributing to the long distance expansion
  truncated to a given order as a function of the distance $r$ (in
  fm).  Middle panel: Short distance cut-off $r_c$ (in fm) dependence
  of the S-wave charmonium spectrum masses (in GeV) for the different
  approximations to the potential. {\bf Right panel:} Relative $ n^3S_1 \to
  e^+ e^-$ decay ratios with respect the leptonic width of the ground state,
  $J/\psi \to e^+e^-$. The bands indicate the corresponding experimental
  number.  In all cases the vertical line marks the critical radius below which
  the full potential becomes imaginary. Solid line: Full BSM. Thick
  dotted: LO approximation. Dotted: NLO. Dotted-dashed: NNLO.}
\label{fig:bsm}
\end{figure*}

Our numerical results are displayed in Fig.~\ref{fig:bsm}. There we
show the BSM potential compared to the linear $V_{\rm LO}(r) = \sigma
r$ (LO), $V_{\rm NLO}(r) = \sigma r + c_1/r$ (NLO) as well as
$V_{\rm NNLO}(r) = \sigma r + c_1/r+ c_3/r^3$ (NNLO)
approximations. Actually, for the typical value of the string tension
the difference between the potentials is at most about $50 {\rm
MeV}$. Applying the renormalization procedure we see that when the
short distance cut-off coincides with the critical distance the
difference among the different masses become much smaller. This is
just a manifestation of the short distance insensitivity embodied by
the renormalization. While in the exact BSM we are forced to stop at
the critical radius, the LO, NLO, NNLO, ... truncations to such a
potential allow to remove the cut-off completely. However, as we see
from the figure this provides in our view an estimate on the
systematic error inherent to the BSM model; the mass shift generated
by the different truncations is about $50{\rm MeV}$. Of course, there
is no point in going beyond such an accuracy before other important
effects (like e.g. relativistic corrections) are taken into account.

Actually, there are hints~\cite{Lambiase:1995st,Nesterenko:1997hw}
that for the Nambu-Goto string the critical radius moves about a factor
of two down (in our case it means $\sim 0.2 {\rm fm }$ ) when quark
masses become comparable to the string tension $m_q / \sqrt{\sigma}
\sim 1$ (in our case $m_c / \sqrt{\sigma} =2.85$)~\footnote{We also note a
recent effective string theory approach~\cite{PerezNadal:2008vm},
where these finite mass effects are evaluated regardless on the
specific string dynamics, yielding a potential, up to a constant term, of
the form 
\begin{eqnarray}
V(r)=\frac{1}{m}\left[\frac{g^{2}\Lambda^{4}}{\pi\kappa}\log(\sqrt{\kappa}\,
r)\right],
\end{eqnarray}
hence they obtain the nontrivial result that the $1/m$ potential must grow
logarithmically at large $r$ and it has to connect with the $1/r^{2}$ behaviour
at short distances of the $1/m$ contribution calculated by perturbation
theory. The comparison of the lattice data is done through the next potential
form
\begin{eqnarray}
V_{\rm fit}(r)= \frac{1}{m} \left[\frac{-c'}{r^2} + d' \ln
\left(\frac{r}{r_0}\right)
\right],
\end{eqnarray}
where once a fit to the lattice data is done~\cite{Koma:2006si,Koma:2006fw}
yields $c'=0.0027 {\rm GeV}^2$ and $d'=0.0075 {\rm GeV}^2$.}

We depict the relative $n^3S_1 \to e^+ e^-$ decay ratios
with respect the ground state $ J/\psi \to e^+ e^-$ decay. As we
have emphasized above the advantage of such a comparison is that purely
perturbative and state independent corrections in the van Royen-Weisskopf
formula are factorized out. However, note that the very use of this formula
assumes a regular potential. Thus we take naturally the wave function at the
shortest cut-off radius, $r_c$. As we can see there are somewhat large
differences for the simple linear potential (LO) approximation as compared to
NLO, NNLO or the full BSM results which agree among them, displaying a
convergent pattern at any value of $r_c$. Differences between NLO and NNLO
start being noticeable below cut-off radius of about $0.15 {\rm fm}$ for the
first excited state, although the value seems slightly displaced to larger
numbers for the next excited states, as one might expect from their stronger
short distance sensitivity. Also interesting is the fully converged result at
$r_c \to 0$, a fact which follows analytically by a detailed study of the short
distance wave functions (see Appendix~\ref{sec:shortdistances}).

These features reinforce the value of renormalization as a way to study the
effect of improving on the long distance components of the $\bar{q}q$
potential. Indeed, even though the NNLO potential strongly diverges at short
distances the effects on the short distance wave function are not out of
control; the relevant scale is provided by the binding energy of the bound
state.

Of course, while the outcoming decay ratios seem to reproduce the experimental
numbers above the critical radius, it is not completely obvious whether
fixing the short distance cut-off radius is fully justified, as the decay
formula corresponds to the strict infinite mass and point-like limit of the
related electromagnetic matrix element $\gamma \to q \bar q $. Actually, the
momentum dependence and recoil effects due to the incoming virtual photon have
been dealt with in Ref.~\cite{Leinweber:1987hi} where the corresponding
modification to the standard decay formula decreases its value substantially. In
appendix \ref{sec:recoil} we collect some relevant formulas. In our case we see
that, as expected, there is an increasing suppression for excited states,
although in the ratio this effect is negligible as can be seen from
Table~\ref{tab:van-royen}.

In summary, the analysis shows that even if one can apply the
renormalization to any long distance approximation to the BSM
potential, due to the existence of the scale $r_0$, one should not
take $r_c \to 0$, but rather to keep $r_c \sim r_0$ to comply with the
dynamics. 

\begin{table}[ttt]
\begin{center}
\begin{tabular}{ccccc}
\hline
\hline
State & & ${\cal R}_{\rm Weisskopf}$ & & ${\cal R}_{\rm Leinweber}$ \\
\hline
$J/\psi$ & & $1.00$ & & $1.00$ \\
$\psi(2S)$ & & $0.58$ & & $0.58$ \\
$\psi(3S)$ & & $0.41$ & & $0.42$ \\
$\psi(4S)$ & & $0.31$ & & $0.32$ \\
\hline
\hline
\end{tabular}
\caption{\label{tab:van-royen} Relative $ n^3S_1 \to e^+ e^-$ decay ratios with
respect the ground state $J/\psi \to e^+ e^-$ decay calculated through the van
Royen-Weisskopf formula and taking into account momentum dependence and recoil
effects according to Ref.~\cite{Leinweber:1987hi}.}
\end{center}
\end{table}

\subsection{Singular repulsive interactions}

To analyze a situation where renormalization conditions cannot be
applied let us consider the case where in the power series expansion
Eq.~(\ref{eq:ipower}) one term happened to be singular and repulsive
at short distances, say for $n=3$ (unlike the BSM discussed above).
Then the solution would be of the form
\begin{equation}
u(r) \to r^{3/4} \left[ C_{1R} \, e^{2\left(\frac{R}{r}\right)^{1/2}} + C_{2R}
\, e^{-2\left(\frac{R}{r}\right)^{1/2}}  \right],
\end{equation}
where we have a converging and a diverging exponential at the origin. 

Clearly, if we impose normalizability of the bound state, the regular
solution must be chosen.  In such a case the bound state energy is
{\it predicted} as it usually happens for the standard non-singular
quantum mechanical potentials.  Conversely, if the bound state energy
is chosen {\it arbitrarily} then the wave function is non-normalizable
at short distances. In this sense, the rigorous results given by
Eq.~(\ref{eq:concavity}), valid in the infinite mass limit, which
imply that {\it all} inverse powers are attractive singularities are
most welcome, because this would imply that the problem of a repulsive
singularity of a {\it truncated} potential is related to the
truncation itself. As we have mentioned, these violations of the
conditions featured by Eq.~(\ref{eq:concavity}) are also observed in
perturbation theory (see e.g.~\cite{Laschka:2011zr}). 

Finally, it is not obvious what might be the situation for
relativistic corrections which on purely dimensional grounds are of
the general form
\begin{eqnarray}
V(r ) \sim \frac{1}{m^n} \frac{p^k}{r^{k+n+1}}  
\end{eqnarray}
Of course, Weyl ordering should be applied to make sure that
the operator is hermitian~\cite{Brambilla:1993zw}. However, as shown in
\cite{Brambilla:2000gk,Brambilla:2001xk} one can eliminate the $p$ dependence in
terms of $L$ dependence at the expense of reshuffling the $1/m$ expansion.
Ultimately, we end up with singular problems which generally depend on the
angular momentum, showing that one must possibly use a different renormalization
condition for each partial wave. However, note that for this to happen one must
have a singular attractive singularity.  This interesting topic deserves further investigation.

\section{Conclusions}
\label{sec:conclusions}

We have reanalyzed the calculation of the $1^{--}$ charmonium spectrum in a
constituent quark model using a renormalization boundary condition scheme. This 
approach avoids explicitly the introduction of phenomenological form factors 
used in the past which are usually fine-tuned to low-lying mesonic observables. 

The renormalization viewpoint actually uses these low-lying properties
as input parameters and sidesteps the fine tuning, as illustrated in
several examples.  For the phenomenologically successful model of
Ref.~\cite{Vijande:2005eu} where {\it ad hoc} form factors are
introduced as regulators, we find an almost perfect agreement with the
renormalization approach.  This result provides confidence on the way
the original model took into account the unknown short distance
dynamics. The use of the renormalization scheme allows us to further
study the dependence of the states on the model parameters in a
cleaner way since the regulator dependence has been removed.

We find that the mass of the excited states strongly depends on the
string tension parameter as one would naturally expect. Further, there
is also a remarkable insensitivity to the strong coupling constant
entering the one gluon exchange contribution of the potential. This
avoids a great deal of unphysical fine tuning which suggested taking
for this parameter unnaturally large values $\alpha_{s}=0.3-0.4$. In
contrast, the leptonic decay widths present a marked dependence on the
strong coupling constant, as expected likewise because these are short
range observables. Of course, a more complete description of
charmonium systems should also take explicitly into account the
contribution to the open charm channels which will provide mass shifts
and widths for unstable quark-antiquark bound states. In summary, the
renormalization approach traces quite naturally and explicitly the
long and short distance dependence of hadronic properties.

\appendix 

\section{Renormalization conditions and Error estimates}
\label{sec:errors}

The boundary condition allows to connect the bound state to the excited states
by the matching condition. Here we estimate what is the error in the excited
mass as a function of the the cut-off radius. This means that we start
with the ground state energy, $E_0$, which is fixed throughout and compute the
logarithmic derivative at $r_c$. Then, we perform a calculation of the
bound state energy, $E_n (r_c)$, which obviously depends on the cut-off radius.
Let us denote $u_n(r,r_c)$ the excited wave function where the dependence on
$r_c$ is explicitly shown. Under an infinitesimal change, $r_c \to r_c +
\Delta r_c$ we get
\begin{eqnarray}
-\Delta u_n'' + U \Delta u_n = \Delta E_n u_n + E_n \Delta u_n,
\end{eqnarray}
Note that the variation is defined for a {\it fixed} value of $r$,  
\begin{eqnarray}
\Delta u_n (r,r_c) = u_n (r,r_c + \Delta r_c)- u_n (r,r_c ),
\end{eqnarray}
and hence 
\begin{eqnarray}
\Delta u_n' (r,r_c) = u_n' (r,r_c + \Delta r_c)- u_n' (r,r_c ),
\end{eqnarray}
where here, the prime denotes derivative with respect to the $r$
variable.  Therefore, if we use the boundary condition
\begin{eqnarray}
u_n' (r_c,r_c) = L_n (r_c) u_n (r_c, r_c),
\end{eqnarray}
we get 
\begin{equation}
\begin{split}
\Delta u_n' (r,r_c) \Big|_{r=r_{c}} =& \, \Delta (L_n (r_c) u_n (r_c, r_c) ) \\
& -
\frac{\partial u_n' (r,r_c)}{\partial r_c} \Big|_{r={r_c}} \Delta r_c.
\end{split}
\end{equation}

From here we obtain 
\begin{eqnarray}
(L_n'+ L_n^2 - U + E_n) u_n = - \frac{\partial u_n'}{\partial r_c} + L_n 
\frac{\partial u_n}{\partial r_c},
\end{eqnarray}
and multiplying this equation by $u_n$ and subtracting the original
equation multiplied by $\Delta u_n$ we get
\begin{eqnarray}
-\Delta u_n'' u_n + u_n'' \Delta u_n  = \Delta E_n u_n^2,
\end{eqnarray}
which integrating from $r_c$ and infinity and using that for a bound state
$u_n (r,r_c) \to 0$ at large distances we get
\begin{eqnarray}
-\Delta u_n' u_n + u_n' \Delta u_n  = \Delta E_n \int_{r_c}^\infty u_n^2.
\end{eqnarray}

On the other hand, since the ground state energy is fixed, $\Delta E_0=0$, we
get
\begin{eqnarray}
-\Delta u_0' u_0 + u_0' \Delta u_0  = 0,
\end{eqnarray}
and taking into account 
\begin{eqnarray}
\Delta E_n = (L_n'+ L_n^2 - U + E_n)
\frac{u_n(r_c)^2}{\int_{r_c}^\infty u_n (r)^2 dr },
\end{eqnarray}
so that using $L_n = L_0$ we arrive  at 
\begin{eqnarray}
\Delta (E_n-E_0)  = (E_n- E_0) \frac{u_n(r_c)^2}{\int_{r_c}^\infty u_n (r)^2
  dr } \Delta r_c.
\end{eqnarray}
Actually, integrating we get 
\begin{eqnarray}
\left[ E_n(r_c)-E_0 \right] \int_{r_c}^\infty \, dr \, u_n (r)^2 = {\rm const.} 
\end{eqnarray}
For a normalized state we have for small $r_c$
\begin{eqnarray}
E_n(r_c)- E_0 =  (E_n - E_0) \left[ 1 + r_c u_n(r_c)^2 + \dots \right] 
\end{eqnarray}

Therefore for a regular potential with a non-trivial boundary
condition $u_n (0) \neq 0$ the error is at least linear. For a
singular and attractive potential, $1/r^n$, the error is ${\cal O}
(r_c^{1+n/2})$ up to some oscillations. As we see, the convergence is
from above and proportional to the energy difference as well. This
means that the effect of putting a finite cut-off fixing the ground
state energy is repulsive and increases with the excitation energy.

\section{Momentum dependence and recoil effects as corrections to leptonic
widths}
\label{sec:recoil}

Momentum dependent effects could reveal significant corrections to the
theoretical leptonic width. The expression for $e^{+}e^{-}$ decay
width of S-wave states in the center-of-mass frame of the meson and
taking into account those effects can be written
as~\cite{Leinweber:1987hi}
\begin{equation}
\begin{split}
&\Gamma(e^{+}e^{-})=16\alpha^{2}e_{q}^{2}\frac{m_{q}^{2}}{M^{4}}\frac{|\vec{k}|
}{E_{e}}(3E_{e}^{2}-|\vec{k}|^{2})\times \\
&
\times\left[\int_{0}^{\infty}\frac{E_{q}+m_{q}}{m_{q}}\frac{|\vec{p}\,|^{2}}{E_{
q}}\left(1+\frac{p^{2}}{3(E_{q}+m_{q})^{2}}\right)\psi(|\vec{p}\,|)d|\vec{p}\,
|\right]^{2},
\label{ref:eqee}
\end{split}
\end{equation}
where $e_{q}$ is the quark charge in units of the charge of the electron, M is the mass of the meson and
\begin{equation}
\begin{split}
&
E_{q}=\left(|\vec{p}\,|^{2}+m_{q}^{2}\right)^{1/2}, \\
&
E_{e}=\left(|\vec{k}|^{2}+m_{e}^{2}\right)^{1/2},
\end{split}
\end{equation}
are the quark and lepton energies respectively. 

The static limit of Eq.~(\ref{ref:eqee}) may be obtained by considering the nonrelativistic limit
\begin{equation}
\frac{|\vec{p}\,|^{2}}{m_{q}^{2}}\ll1
\end{equation}
and in this case the integral in Eq.~(\ref{ref:eqee}) reduces to
\begin{equation}
\frac{2}{m_{q}M^{2}}\int_{0}^{\infty}\psi(|\vec{p}\,|)|\vec{p}\,|^{2}d|\vec{p}\,|.
\label{ref:eqI}
\end{equation}

Recalling the Fourier transform
\begin{equation}
\phi(\vec{r})=\frac{1}{(2\pi)^{3/2}}\int e^{i\vec{p}\cdot\vec{r}}\psi(\vec{p}\,)d\vec{p}
\end{equation}
and evaluating at the origin, we have
\begin{equation}
\phi(\vec{r}=0)=\frac{1}{(2\pi)^{3/2}}\int\psi(\vec{p}\,)d\vec{p}=\sqrt{\frac{2}{\pi}}\int_{0}^{\infty}|\vec{p}\,|^{2}\psi(|\vec{p}\,|)d|\vec{p}\,|,
\end{equation}
so the integral above, Eq.~(\ref{ref:eqI}), becomes
\begin{equation}
\frac{2}{m_{q}M^{2}}\int_{0}^{\infty}\psi(|\vec{p}\,|)|\vec{p}\,|^{2}d|\vec{p}\,|=\frac{(2\pi)^{1/2}}{m_{q}M^{2}}|\phi(\vec{r}=0)|^{2}.
\end{equation}

With the kinematic relationships, the static limit of Eq.~(\ref{ref:eqee}) is
\begin{equation}
\Gamma(e^{+}e^{-})\Big|_{static}=\frac{16\pi\alpha^{2}Q^{2}}{M^{2}}|\phi(\vec{r}
=0)|^{2},
\end{equation}
which is the well-known van Royen-Weisskopf formula.

\begin{figure*}[ttt]
\begin{center}
\epsfig{figure=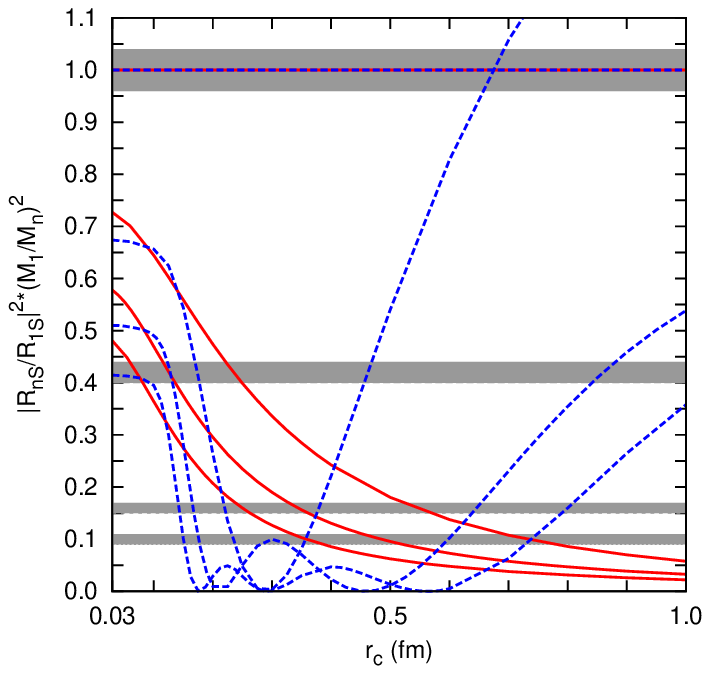,height=5cm,width=5.5cm}
\epsfig{figure=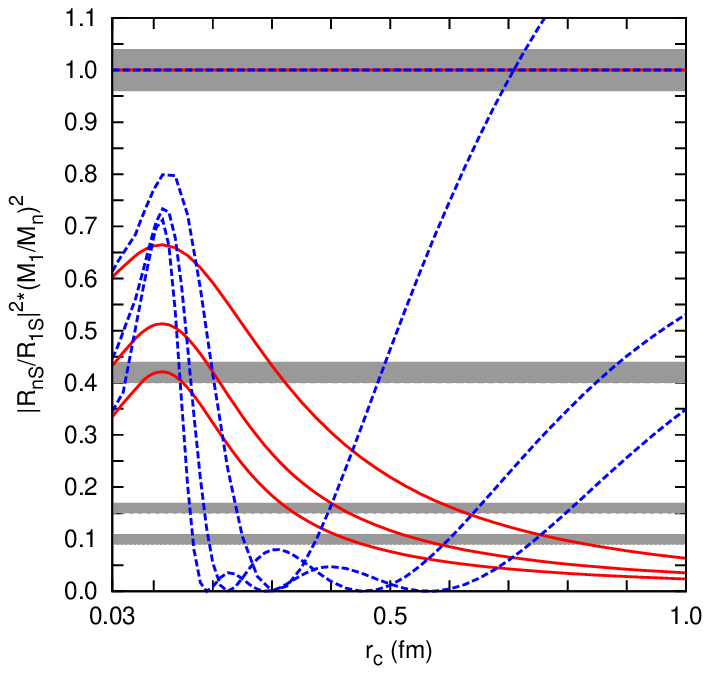,height=5cm,width=5.5cm}
\epsfig{figure=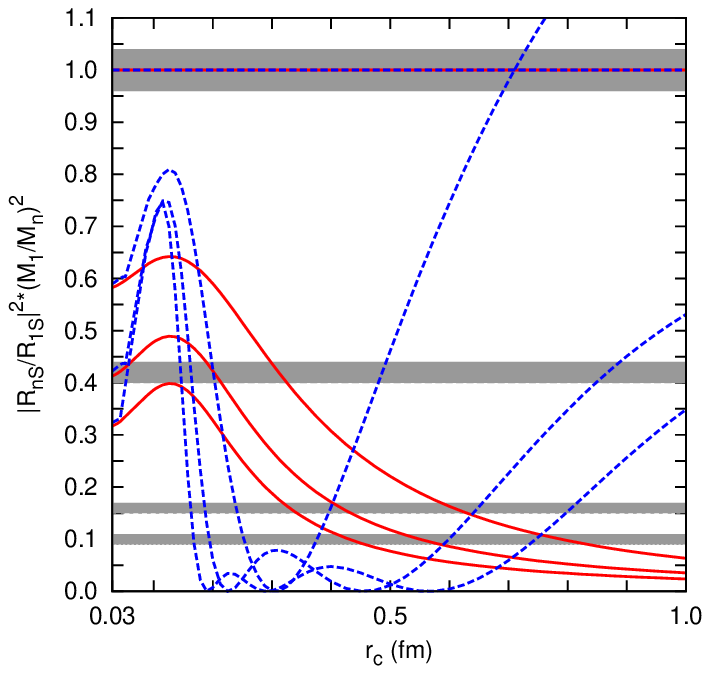,height=5cm,width=5.5cm}
\end{center}
\caption{Relative $n^{3}S_{1}\to e^{+}e^{-}$ decay ratios with respect the
  leptonic width of the ground state, $J/\psi \to e^+e^-$ which are
  calculated by van Royen-Weisskopf (solid line) and Leinweber (dashed line)
  formulas. {\bf Left panel:}  LO. {\bf Middle panel:} NLO. {\bf Right panel:}
  NNLO.}
\label{fig:leinweber}
\end{figure*}

Figure~\ref{fig:leinweber} shows the comparison of the relative
$n^{3}S_{1}\to e^{+}e^{-}$ decay ratios with respect the leptonic width of the
ground state, $J/\psi \to e^+e^-$. The ratios are calculated by van
Royen-Weisskopf and Leinweber formulas. Left, middle and right panels of the
figure show the results for the different terms contributing to the long
distance expansion of the bosonic string model truncated to a given order (LO,
NLO and NNLO).

\section{Short distance analysis of wave functions}
\label{sec:shortdistances}

\subsection{Single channel case}
\label{sec:shortdistances_single}

In this appendix we summarize a few interesting facts concerning the
short distance behaviour of wave functions. In the OGE case an
attractive Coulomb like behaviour holds, so that at short distances the reduced 
potential reads
\begin{eqnarray}
U(r) \equiv 2 \mu V(r)  \to -\frac{1}{R r},
\end{eqnarray}
so that for the nS state we have short distance behaviour 
\begin{eqnarray}
u_n(r) \to A_n \left[ 1 - \frac{3r}{2R}- \frac{r}{R} \log \left(
  \frac{r}{R} \right) \right] + B_n r,
\end{eqnarray}
which is a linear combination of the regular wave function and the
singular one. Note that since the ground state has a given energy the
irregular component does not vanish. On the other hand, the
orthogonality condition,  Eq.~(\ref{eq:bc}), implies 
$A_n/B_n=A_0/B_0$ so that 
\begin{eqnarray}
\frac{u_n(r_c)}{u_0(r_c)} \to \frac{A_n}{A_0},
 \end{eqnarray}
which shows that the ratio between wave functions becomes finite as
the cut-off is removed, as can be seen at the right panel in
Fig.~\ref{fig:bsm}.

For a power-like short distance singular potential we may keep the
strongest singularity
\begin{eqnarray}
V(r) \to -\frac{C_n}{r^n}.
\label{eq:sing_U}
\end{eqnarray}

The solution of the Schr\"odinger equation requires the reduced potential 
\begin{eqnarray}
U(r) \equiv 2 \mu V(r)  \to -\frac{1}{R^2} \left(\frac{R}{r} \right)^n,
\end{eqnarray}
where for convenience the variable $R = (2\mu C_n)^{1/4}$ with length
scale dimensions has been introduced. At short distances the reduced
de Broglie wavelength is given by
\begin{eqnarray}
\lambda (r) \equiv \frac{1}{\sqrt{U(r)}} = R \left(\frac{r}{R}
\right)^{\frac{n}{2}},
\end{eqnarray}
which fulfills 
\begin{eqnarray}
\frac{d \lambda (r)}{d r} \frac{n}{2} \left(\frac{r}{R}
\right)^{\frac{n}{2}-1} \ll 1,
\end{eqnarray}
for $r \ll R$. In such a case the WKB method can be
applied~\cite{Frank:1971xx} yielding
\begin{eqnarray}
u(r) &&\to u_{\rm WKB} (r)= \nonumber \\ &&= \frac{A}{[-U(r)]^{1/4}}
\sin \left[ \int dr \sqrt{-U(r)}+ \varphi \right],
 \end{eqnarray}
where $A$ and $\varphi$ are undetermined amplitude and phase which may be
obtained by matching to the exact solution in the region $r \sim
R$. In the case of the singular potential given by
Eq.~(\ref{eq:sing_U}) we have for the m-state 
\begin{equation}
u_m(r)\to A_m \left(\frac{r}{R} \right)^{\frac{n}{4}}
\sin \left[ \frac{2}{n-2}
\left(\frac{r}{R} \right)^{\frac{n}{2}-1}
+ \varphi_m \right].
\end{equation}

However, the orthogonality condition, Eq.~(\ref{eq:bc}), imply
$\varphi_m=\varphi_0$. Thus we obtain 
\begin{eqnarray}
\frac{u_m(r_c)}{u_0(r_c)} \to \frac{A_m}{A_0},
 \end{eqnarray}
which shows that the ratio between wave functions becomes finite as
the cut-off is removed. For $n=3$ this is seen at the right panel in
Fig.~\ref{fig:bsm}.

\subsection{Coupled channel case}
\label{sec:shortdistances_coupled}

Here we undertake the short distance analysis of the spin-orbit and tensor
interactions. At short distances one may neglect all terms and just keep the
$1/r^3$ singular contribution yielding
\begin{eqnarray}
\begin{pmatrix} -u''(r) \\ -w''(r) \end{pmatrix}
\pm \frac{R}{r^3}\begin{pmatrix} 0 & \frac{2\sqrt{2}}{3}
 \\ \frac{2\sqrt{2}}{3}  & -\frac{20}{3}  \end{pmatrix} \begin{pmatrix} u(r)
  \\ w (r) \end{pmatrix} =0 \, . \nonumber \\ 
\label{eq:sch_coupled-dip} 
\end{eqnarray}

This system can be diagonalized by going to the rotated basis 
\begin{eqnarray}
\begin{pmatrix} v_{1}(r)
  \\ v_{2} (r) \end{pmatrix}
 &=& \begin{pmatrix}  \cos \alpha  &  \sin \alpha  \\ 
-\sin \alpha &  \cos \alpha 
\end{pmatrix} \, \begin{pmatrix} u(r)
  \\ w(r) \end{pmatrix},
\end{eqnarray} 
where the new functions satisfy 
\begin{eqnarray}
-v_{1} '' (r) + \frac{R_1}{r^3} v_{1} (r) &=& 0, \nonumber \\ 
-v_{2} '' (r) - \frac{R_2}{r^3} v_{2} (r) &=& 0,
\label{eq:decoup}
\end{eqnarray}
and the $R_1$ and $-R_2$ are the corresponding eigenvalues 
\begin{eqnarray}
R_1 &=& -\frac{10}{3} + 2\sqrt{3} > 0, \nonumber \\ 
-R_2 &=& -\frac{10}{3} - 2\sqrt{3} < 0,
\end{eqnarray}
and the mixing angle is $\theta = 1.1^0$, a rather small value. At short
distances the solutions of Eq.~(\ref{eq:decoup}) could be analyzed via the WKB
method as we have done in the previous section, but for
this case we can undertake the analysis in terms of Bessel functions,
which short distance is analytically known. Actually, the solutions of
$ -y''(x) - y(x)/x^3=0$ are
\begin{eqnarray}
\sqrt{x} J_1 ( 2 /\sqrt{x}) &=& - \frac{x^\frac34}{\sqrt{\pi}} 
\cos(\pi/4+2 /\sqrt{x}) + \dots,  
\nonumber \\
\sqrt{x} Y_1 ( 2 /\sqrt{x}) &=& - \frac{x^\frac34}{\sqrt{\pi}} 
\cos(\pi/4-2 /\sqrt{x}) + \dots, \nonumber \\
\end{eqnarray}
whereas the  solutions of $ -y''(x) + y(x)/x^3=0$ are
\begin{eqnarray}
\sqrt{x} K_1 ( 2 /\sqrt{x}) &=& \frac12 \sqrt{\pi }x^\frac34 e^{-2/\sqrt{x}}
+ \dots, \nonumber \\
\sqrt{x} I_1 ( 2 /\sqrt{x}) &=& \frac1{2 \sqrt{\pi }} x^\frac34 e^{2/\sqrt{x}}
+ \dots, 
\end{eqnarray}

All this amounts to write the solutions in the suitable form
\begin{eqnarray}
v_{1} (r) &\to & \left(\frac{r}{R_1}\right)^{\frac34} \left[ C_{1R} 
e^{+2 \sqrt{\frac{ R_1}{r}}} + C_{2R} e^{- 2\sqrt{\frac{ R_1}{r}}} \right],
\nonumber \\
v_{2} (r) &\to
& C_A \, \left(\frac{r}{R_2}\right)^{\frac34} 
\sin \left[ 2 \sqrt{\frac{ R_2}{r}} +  \varphi \right].
\label{eq:short_bc_r3}
\end{eqnarray}

The four constants appearing here, $C_{1R}$, $C_{2R}$, $C_{A}$ and $\varphi$
reflect that the total order of the system is four. The last equation also shows
that generally solutions will diverge as $e^{2 \sqrt{R_1/r}}$ at the origin,
hence preventing the bound state normalization condition, unless $C_{1R}=0$. In
such a case the normalizable solution may be written as
\begin{equation}
\begin{split}
&
\begin{split}
u(r) \to & \cos \alpha \, C_A \, \left(\frac{r}{R_2}\right)^{\frac34} 
\sin \left[ 2 \sqrt{\frac{ R_2}{r}} +  \varphi \right] \\
&
-\sin\alpha \, C_{2R} \, \left(\frac{r}{R_2}\right)^{\frac34} e^{-2 \sqrt{\frac{
R_2}{r}}},
\end{split} \\
&
\\
&
\begin{split}
w(r) \to & \sin \alpha \, C_A \, \left(\frac{r}{R_2}\right)^{\frac34} 
\sin \left[ 2 \sqrt{\frac{ R_2}{r}} +  \varphi \right] \\
&
+\cos \alpha \, C_{2R} \, \left(\frac{r}{R_2}\right)^{\frac34} e^{-2
\sqrt{\frac{ R_2}{r}}}.
\end{split}
\end{split}
\end{equation}

The three independent constants appearing here for the regular solution $C_A$,
$C_{2R}$ and $\varphi$ correspond to fix the energy $M_{J/\psi}$, the asymptotic
$D/S$ ratio $\eta$, and the normalization condition. Note that when integrating
from large distances to short distances with a given bound state energy and an
arbitrary $\eta$ we would always have a contribution from the exponentially
diverging solution since $C_{1R} \neq 0$. Thus, the condition $C_{1R}=0$
{\it predicts} $\eta$ from the bound state energy. The foregoing analysis shows
that for the $1/r^3$ singularity appearing here the solution is ambiguous and
the bound state energy for a given state has to be treated as an {\it input}. Of
course, the orthogonality requirement between different states implies that if
one state, $(u_n,w_n)$, has a short distance phase $\varphi_n$ and another
state, $(u_m,w_m)$, has a short distance phase $\varphi_m$, one has
\begin{eqnarray}
0 &=& 2 \mu (M_n - M_m)  \int_0^\infty \, dr \, (u_n u_m + w_n w_m) \nonumber \\
&=& 
\left[u_{n}' u_{m}  - u_{n} u_{m}'  + w_{n}' w_{m}  - w_{n} w_{m}' \right]\Big|_0^{\infty} \nonumber \\ 
&=& \frac1{R_2} C_{A,n} C_{A,m} 
\sin ( \varphi_n - \varphi_m) 
\end{eqnarray}
whence $\varphi_n=\varphi_m$ is obtained. This shows that all states are linked
through the spin-dependent splitting provided the ground state energy is given.

\section{Fitting parameters of the renormalized model}
\label{sec:fitting}

It is interesting to see what is the renormalized model sensitivity to
the parameters. Of course, at this level one should envisage the
possibly non-negligible contribution of the missing mesonic thresholds
and the induced mass shift.  To this end we fit the model parameters
from the experimental masses of $1^{--}$ $c\bar{c}$
states~\cite{Nakamura:2010zzi} and estimate their theoretical
uncertainties as well as an educated mass shift guess.  The way to
perform the first task is of course debatable.  This requires some
compromise as to what are the errors attached to the masses within the
present framework. Naively one would just take the quoted PDG errors
on the experimental masses. However, as we have mentioned above, the
present model incorporates thresholds but no explicit coupling to
meson-meson channels. This affects more significantly
the higher excited states via a subthreshold induced mass shift, which
we take as a systematic error of the model.

We decide to make a $\chi^2$ fit using as fitting parameters the quark mass,
$m_c$, $a_c$ and $\mu_c$ related with the confinement strength and the strong
coupling constant, $\alpha_s$. We define the
\begin{equation}
\chi^2(\vec p) = \sum_i \frac{(M_{exp}(i)-M_{teo}(i,\vec
p))^2}{\sigma_{exp}(i)^2},
\end{equation}
where $\vec p$ represents our model parameters and the experimental data are taken as
$M_{\psi(2S)} = 3686.093 \pm 0.143 \, {\rm MeV}$, $M_{\psi(3S)} =
4039.6 \pm 42.25 \, {\rm MeV}$ and $M_{\psi(4S)} = 4361 \pm 37 \, {\rm MeV}$,
where as explain previously the errors are taken as the half-width of the state. 
By minimizing the $\chi^2$ function we
obtain the theoretical uncertainties from the corresponding covariance matrix at
the minimum. The outcoming values for the parameters are
\begin{eqnarray}
m_c &=& 1862\pm12.6 \, {\rm MeV} \,\,\, (0.68\,\%),
\nonumber \\
a_c &=& 524  \pm 43  \, {\rm MeV}   \,\,\,\,\,\,\,\,\,\,\, (8.2 \,\%),
\nonumber \\
\mu_c &=& 88\pm7.2 \, {\rm MeV} \,\,\,\,\,\,\,\,\, (8.2 \,\%),
\nonumber \\
\alpha_s &=& 0.41\pm0.14       \,\,\,\,\,\,\,\,\,\,\,\, (34\,\%),
\end{eqnarray}
where we put in parenthesis the relative uncertainties in percentage.
Here we clearly see that the highest uncertainty corresponds to the
value of the strong coupling constant. This rather large insensitivity
to the otherwise too large $\alpha_s$ is a rewarding feature of the
renormalization approach. Actually, quite natural values of $\alpha_s$
are obtained. The spectrum at the $\chi^2-$minimum is given by
\begin{eqnarray}
\psi(2S) &=& 3687 \pm 80  \, {\rm MeV},
\nonumber \\
\psi(3S) &=& 4108 \pm 79  \, {\rm MeV},
\nonumber \\
\psi(4S) &=& 4348 \pm 80  \, {\rm MeV},
\nonumber \\
\psi(5S) &=& 4586 \pm 66  \, {\rm MeV},
\end{eqnarray}



\bibliography{ren-charm}

\end{document}